\documentclass[aps,amsmath,twocolumn,amssymb,titlepage,superscriptaddress,10pt,longbibliography]{revtex4-1}
\usepackage[T1]{fontenc}
\usepackage[utf8]{inputenc}
\usepackage{amsmath}
\usepackage{braket}
\usepackage{amsfonts}
\usepackage{comment}
\usepackage{graphicx}
\usepackage{hyperref}
\usepackage[capitalize]{cleveref}
\usepackage{amssymb}
\usepackage{amsmath}
\usepackage{mathtools}
\usepackage{todonotes}

\DeclareMathOperator{\Tr}{Tr}

\newcommand{\Ham}{H}

\newcommand{\bbC}{{\mathbb C}}

\newcommand{\bbZ}{{\mathbb Z}}
\newcommand{\caB}{{\mathcal B}}

\definecolor{darkgreen}{rgb}{0,0.7,0}

\begin{document}
\title{Symmetry breaking and thermal  phase transition of the spin-1 quantum magnet\\ with SU(3) symmetry on the simple cubic lattice}
\author{Nils Caci}
\affiliation{Laboratoire Kastler Brossel, Coll\`ege de France, CNRS, \'Ecole Normale Sup\'erieure - Universit\'e PSL, Sorbonne Universit\'e, 75005 Paris, France}
\author{Dominik Chudy}
\affiliation{Institute for Theoretical Solid State Physics, JARA FIT and JARA CSD, RWTH Aachen University, 52056 Aachen, Germany}
\author{Pablo Daniel Mendez Mariscal}
\affiliation{Institute for Theoretical Solid State Physics, JARA FIT and JARA CSD, RWTH Aachen University, 52056 Aachen, Germany}
\author{Daniel Ueltschi}
\affiliation{Department of Mathematics, University of Warwick, Coventry CV4 7AL, United Kingdom}
\author{Stefan Wessel}
\affiliation{Institute for Theoretical Solid State Physics, JARA FIT and JARA CSD, RWTH Aachen University, 52056 Aachen, Germany}

\begin{abstract}
Using a combined analysis from Poisson-Dirichlet and symmetry-breaking calculations as well as quantum Monte Carlo simulations, we examine the ordered phase and the thermal phase transition of the three-dimensional spin-1 quantum magnet on the simple cubic lattice with bilinear and biquadratic interactions and SU(3) internal symmetry. We obtain exact results for the  order parameter distribution function that compare well to the quantum Monte Carlo data. Furthermore, based on a detailed finite-size analysis, we provide evidence that the thermal melting transition at the SU(3) point is either weakly first-order in this system or, if continuous, it falls beyond the unitary-bounds of conformal field theory. 
\end{abstract}

\maketitle
\section{Introduction}\label{Sec:Intro}
Phase transitions in quantum many-body systems are a central focus of modern condensed matter physics. In recent years, particular attention has been given to quantum systems that exhibit phase transitions with subtle characteristics, such as unconventional critical scaling or an unexpectedly large but finite correlation length at the  transition point --  referred to as weakly first-order transitions~\cite{Kaplan2009,Nahum2015,Wang2017,Gorbenko2018,Gorbenko2018II,Ma2019,Nogueira2019,Iino2019,Demidio2021,Caci2023}.
A rather versatile opportunity for studying such phenomena is provided by the field of quantum magnetism. One central aspect in both the exploration and the characterization of phase transitions in such systems is the symmetry of the Hamiltonian that governs the dynamics of the quantum spins. 
In many realistic models of quantum magnetic materials, the underlying crystal and orbital structure leads to exchange interactions with a reduced symmetry, i.e., forming a proper subgroup of the spin SU(2) symmetry of the bilinear Heisenberg spin exchange interaction. 

There are however also model systems,  in which the actual internal symmetry of the Hamiltonian is {\it larger} than the SU(2) symmetry of the basic bilinear Heisenberg spin exchange. An important  example from basic quantum magnetism, for which such an enhanced symmetry is realized, is the spin-1 model with both bilinear and biquadratic exchange interactions, tuned to  equal strength, described by the Hamiltonian

\begin{equation}
\label{Eq:ham}
    \Ham = -\: J \!\!\sum_{\langle i,j\rangle \in\caB_\Lambda}\!\!\left[   \,( \mathbf{S}_i \cdot \mathbf{S}_j )+   \, (\mathbf{S}_i \cdot \mathbf{S}_j)^2\: \right].
\end{equation}
Here, $\caB_\Lambda$ denotes a set of (usually only nearest-neighbor) bonds  of the underlying lattice $\Lambda$  and we consider  the case $J>0$. In the past, the quantum phase diagram of the generic version of this model, in which the strengths of the bilinear and biquadratic terms are varied individually, has been examined for various lattices in different dimensions, exhibiting in general a rich phenomenology in terms of ground states and quantum phase transitions: Besides conventional ferro- and antiferromagnetic regimes, on a simple cubic lattice also spin nematic phases with long-ranged quadrupolar correlations are stabilized at finite temperatures~\cite{Tanaka_2001,Harada2002,Batista2004,Toth2012,Fridman2013, Ueltschi2015,Caci2023}. While a previous quantum Monte Carlo (QMC) study of this system concluded that the thermal phase transition out of the ferroquadrupolar spin nematic phase is continuous~\cite{Harada2002}, it was recently demonstrated that the thermal melting of this spin nematic order is in fact weakly first-order, based on the analysis of finite-size QMC data on length scales beyond those previously accessible for this SU(2) symmetric model~\cite{Caci2023}.

Here, we extend these latter investigations  of the bilinear biquadratic spin-1 model on the simple cubic lattice to the case of equal coupling strengths, i.e., we consider the Hamiltonian $H$ from Eq.~(\ref{Eq:ham}). In this case, the global symmetry is in fact enhanced to SU(3). This  becomes transparent upon expressing  $H$ in terms of transposition operators. Namely, denoting by $T_{ij}$ the operator that swaps local states on sites  $i,j\in \Lambda$, i.e.,  $T_{ij} (|\psi\rangle_i\otimes |\chi\rangle_j)=|\chi\rangle_i\otimes |\psi\rangle_j$, 
the Hamiltonian $H$ can be written as
\begin{equation}
\label{Eq:hamP}
    \Ham = -\: J \!\!\sum_{\langle i,j\rangle \in\caB_\Lambda}\!\! \left[1 + T_{ij}\right],
\end{equation}
which makes the SU(3) symmetry of $H$ explicit (as a global transformation $U_\Lambda=\otimes_{i\in\Lambda} U_i$,
with a uniform local unitary $U_i=U$ leaves all $T_{ij}$
invariant, $U_\Lambda T_{ij}U_\Lambda^{-1}=T_{ij}$). Due to the enhanced symmetry of the Hamiltonian,  the low-temperature ferromagnetically ordered phase is characterized by {\it extremal Gibbs states} labeled by
rank-1 projections on $\mathbb{C}^3$, or, equivalently, by the complex projective space $\mathbb{CP}^2$ (i.e., the set of equivalence classes of vectors in $\mathbb{C}^3$  differing only by finite complex factors): This $\mathbb{CP}^2$-description  of the symmetry-broken phase at the SU(3) symmetric point was previously discussed for the Hamiltonian $H$ on a complete graph (for which $\caB_\Lambda$ contains all the bonds between all pairs of sites of $\Lambda$) \cite{Bjornberg2020}, corresponding to the mean-field system. Below, we demonstrate that such a characterization can also be derived for the more physical case of a simple cubic lattice. In particular, we show that this approach can be used to calculate the order parameter distribution function in the SU(3) symmetry broken phase. From this distribution,  one can then also calculate moments of the order parameter distribution as well as cumulant ratios, such as the Binder ratio. 

An alternative description of the Hamiltonian $H$ in terms of a random loop model has also been discussed previously \cite{Toth1993, Aizenman1994, Ueltschi2013}, and allows for a similarly exact calculation of the order parameter distribution function and cumulant ratios based on the Poisson-Dirichlet distribution (PD) conjecture \cite{Goldschmidt2011, Ueltschi2013, Grosskinsky2012, Nahum2013, Ueltschi2017, Caci2023}, which we summarize for the specific model at hand  further below.
Finally, we use QMC simulations to compare these analytical results to unbiased numerical data, and report excellent agreement for both the order parameter distribution function as well as the
 cumulant ratios. 

Finally, we study the thermal melting transition from the ordered phase of the SU(3) symmetric model, based on large-scale QMC simulations. In previous work, it was concluded that this transition is continuous, with critical exponents similar to those for the melting of the spin nematic phase (which was also concluded in that reference to be continuous)~\cite{Harada2002}. Indeed, 
as detailed below, we find that in contrast to the case of the spin nematic regime, where clear signatures of weakly first-order behavior can be obtained from studying  histograms of appropriate observables, such as the energy, at the SU(3) symmetric point~\cite{Caci2023}, no such clear indication for first-order behavior emerges. However, we obtain indirect evidence in favor of first-order behavior from using a recently proposed simulation protocol that is based on the application of an explicit symmetry-breaking field of strength $\lambda$ that couples to the order-parameter at the transition temperature~\cite{Demidio2021}. One then monitors the flow of a running exponent $1/\delta(L, \lambda)$, which for a continuous transition approaches the inverse of the actual critical exponent $1/\delta$ for linear system size $L\rightarrow\infty$ and $\lambda\rightarrow 0$. Our analysis shows that if the thermal melting would  be continuous, then the value of $1/\delta$ would fall below the unitarity bound~\cite{Rychkov2017} of 1/5 from conformal field theory (CFT). While we cannot exclude that the thermal transition has a continuous
non-unitary CFT character, a weakly first-order scenario appears to be the more natural  explanation for our numerical findings. We note that a recent study of the three-dimensional (3D) CP${}^2$ models, reports clear evidence for a weakly first-order thermal phase transition as well~\cite{Pelissetto2019}, and this model emerges as an effective theory of certain SU(3) quantum antiferromagnets~\cite{Read1990}. 

The remainder of this work is organized as follows: In Sec.~\ref{Sec:Methods} we discuss  the symmetry breaking calculations and  the PD conjecture approach for the SU(3) symmetric model, as well as the QMC method that we used for this study. Our results for the order parameter distribution and the cumulant ratios are provided in Sec.~\ref{Sec:OP}, while the QMC results for the thermal melting transition are presented in Sec.~\ref{Sec:TM}. Furthermore, we document benchmark Monte Carlo applications of the method from Ref.~\cite{Demidio2021} to the classical Potts model in two dimensions (2D), 3D biquadratic XY model, as well as the melting of the 3D spin nematic phase of the bilinear-biquadratic spin-1 model in App.~\ref{Sec:Potts},~\ref{Sec:XY}, and~\ref{Sec:Nematic}, respectively. 
All the numerical data of this study is available online \cite{Caci2025_data}.

\section{Methods}\label{Sec:Methods}

In the following, we first review the symmetry breaking calculations and then discuss the PD conjecture approach in order to analytically obtain the order parameter distribution function within the SU(3)-symmetry broken regime of the Hamiltonian $H$. As already mentioned, an appropriate order parameter is based on local operators $M_i = P_i - \frac13$, with $P_i=P$ equal to an arbitrary rank-1 projector ($i\in\Lambda)$, where we already subtracted the non-zero high-temperature value $\frac13$ of the expectation value of $P_i$, cf. Sec.~\ref{Sec:SB}.
A convenient choice is the projector $|0\rangle\langle 0|_i$ on the $m=0$ state $|0\rangle_i$, for which $M_i$ then equals the negative of the nematic order parameter $Q_i=(S_i^{\rm z})^2 - \frac23$. Below, we calculate the characteristic function $\langle \exp \bigl\{ \frac{{\rm i}k}{|\Lambda|} \sum_{i \in \Lambda} M_i \bigr\} \rangle$  (here, $|\Lambda|$ denotes the number of lattice sites), which is related to the order parameter distribution function
\begin{equation}
\rho(s)=\Bigl\langle \delta \Bigl( s-\frac{1}{|\Lambda|} \sum_{i \in \Lambda} M_i \Bigr) \Bigr\rangle 
\end{equation}
itself by
\begin{equation}
    \Bigl\langle \exp \Bigl\{ \frac {{\rm i}k}{|\Lambda|} \sum_{i \in \Lambda} M_i \Bigr\} \Bigr\rangle = \int_{-\infty}^\infty \rho(s) {\rm e}^{{\rm i}ks} {\rm d}s,
\end{equation}
and from which  $\rho$ follows via inverse Fourier transformation. In the above, the expectation values $\langle \cdot \rangle $ are taken with respect to  the symmetric Gibbs state that is obtained by an appropriate  integral over the extremal Gibbs states, as detailed in Sec.~\ref{Sec:SB}, and which is directly accessible by QMC simulations. Indeed, 
as we will show in Sec.~\ref{Sec:OP}, our analytical prediction for $\rho$ compares well to unbiased QMC simulations for the SU(3)-symmetric model described by the Hamiltonian $H$. 

\subsection{Symmetry breaking calculations}\label{Sec:SB}

As is well-known, the set of Gibbs states is convex and any Gibbs state can be written as an average over extremal ones \cite{Ruelle1969, Israel_2016}. Although knowledge of the extremal Gibbs states provides valuable information, they are not always easy to identify. The symmetric Gibbs state is often much easier to access, since it is the limit of finite-volume Gibbs states. An important question is how to extract information about the extremal Gibbs states from the symmetric one. In this subsection we make an Ansatz regarding the structure of extremal states, and we use it to compute the order parameter characteristic function. This calculation will be repeated using a different heuristics in the next subsection, which will indeed confirm the Ansatz.

In order to illustrate the relation between the nature of extremal and symmetric Gibbs states as well as the order parameter characteristic function in a more familiar context, we first consider the  $\mathbb{Z}_2$-symmetric ferromagnetic Ising model. When the dimension $d\geq2$ and the temperature is low, it has two translation-invariant extremal Gibbs states: $\langle \cdot \rangle^{\rm extr}_+$ and $\langle \cdot \rangle^{\rm extr}_-$, which can be obtained as infinite-volume limits of states with $+$ and $-$ boundary conditions (with $+,-$ denoting the two orientations of the Ising variables), or with the help of magnetic fields in the $+$ and $-$ direction --- the latter construction also works for  quantum systems. The symmetric Gibbs state is then equal to  $\langle \cdot \rangle = \frac12 \langle\cdot\rangle^{\rm extr}_+ + \frac12 \langle\cdot\rangle^{\rm extr}_-$. The extremal states are "clustering" (they have short-range correlations) and it follows that (as $\Lambda \to \bbZ^d$) $ \bigl\langle \exp \bigl\{ \frac {{\rm i}k}{|\Lambda|} \sum_i M_i \bigr\} \bigr\rangle^{\rm extr}_\pm = {\rm e}^{\pm {\rm i} k m_{\rm I}(\beta)}$, where $m_{\rm I}(\beta)$ is the spontaneous magnetization, with $\beta=1/T$ the inverse temperature (we set $k_B=1$). The Ising order parameter characteristic function is then 
\begin{equation}
\begin{split}
     \Bigl\langle \exp \Bigl\{ \frac {{\rm i}k}{|\Lambda|} \sum_{i \in \Lambda} M_i \Bigr\} \Bigr\rangle &= \tfrac12 {\rm e}^{+{\rm i} k m_{\rm I}(\beta)} + \tfrac12 {\rm e}^{-{\rm i} k m_{\rm I}(\beta)} \\
     &= \cos\bigl( k\:  m_{\rm I}(\beta) \bigr).
\end{split}
\end{equation}
The goal of this section is to apply these ideas to the quantum spin-1 model with SU(3) symmetric interactions.

We describe spin-1 systems but the approach works for general spin systems described by transposition operator Hamiltonians (the spin-$1/2$ case corresponds to the ferromagnetic Heisenberg model). The discussion below follows the one in Ref.~\cite{Bjornberg2020}, which deals with mean-field systems.
We assume that the (translation-invariant) extremal Gibbs states can be defined with the help of the rank-1 projectors. Namely, with $P_i=P$ a uniform rank-1 projector on the local Hilbert space, we define
\begin{equation}
\label{extremal states}
    \langle \cdot \rangle^{\rm extr}_P = \lim_{h \to 0+} \lim_{\Lambda \to \mathbb Z^d} \langle \cdot \rangle_{H - h\sum_{i\in\Lambda} P_i}.
\end{equation}
Here, we added an ``external field" term $\sum_i P_i$ to the Hamiltonian, took the thermodynamic  limit, and let the field strength $h$ tend to 0. At high temperatures (or at any positive temperature when $d=1,2$) the Gibbs state is unique and we have $\langle P_i \rangle^{\rm extr}_P = \frac13$ by symmetry. We thus define the order parameter to be
\begin{equation}
\label{def m}
    m(\beta) = \Bigl\langle  \frac{1}{|\Lambda|} \sum_{i \in \Lambda} M_i \Bigr\rangle^{\rm extr}_P = \langle P_{i=0} \rangle^{\rm extr}_P-\tfrac13,
\end{equation}
where $M_i = P_i - \frac13$ and the choice of the lattice site. 
It should be emphasized that this symmetry breaking conjecture is not obvious.
But it is confirmed by mean-field calculations (see \cite{Bjornberg2020}) and
by the PD calculation approach as well as our QMC simulations (see the
following sections).
In the following, we thus consider $i=0$ and for simplicity omit the index $i$.

The Gibbs state $\langle\cdot\rangle$ with SU(3) symmetry has therefore the decomposition
\begin{equation}
    \langle \cdot \rangle = \int \langle \cdot \rangle^{\rm extr}_P \, {\rm d} P,
\end{equation}
where ${\rm d}P$ is the uniform probability measure over rank-1 projectors. It is convenient to integrate over the space $\mathcal U(3)$ of unitary matrices on $\bbC^3$. Namely, 
\begin{equation}
    \langle \cdot \rangle = \int_{\mathcal U(3)} \langle \cdot \rangle^{\rm extr}_{UPU^{-1}} \, {\rm d}U = \int_{\mathcal U(3)} \langle U^{-1} (\cdot) U\rangle^{\rm extr}_P \, {\rm d}U.
\end{equation}
Here, we identified the matrix $U$ on $\bbC^3$ with $U_{i=0}$ appearing in the uniform tensor product $U_\Lambda = \otimes_{i\in\Lambda} U_i$.
In the limit of infinite volumes, we thus have
\begin{equation}
\begin{split}
    \Bigl\langle \exp \Bigl\{ \frac h{|\Lambda|} \sum_{i \in \Lambda} M_i \Bigr\} \Bigr\rangle \! &= \! \int_{\mathcal U(3)} \! \Bigl\langle \exp \Bigl\{ \frac h{|\Lambda|} \sum_{i \in \Lambda} U^{-1} M_i U \Bigr\} \Bigr\rangle^{\rm extr}_P {\rm d}U \\
    &= \int_{\mathcal U(3)} {\rm e}^{h \langle U^{-1} M_{i=0} U \rangle^{\rm extr}_P} \, {\rm d}U.
\end{split}
\end{equation}
The second identity holds because extremal Gibbs states have short-range correlations and $\langle \cdot \rangle^{\rm extr}_P$ is translation-invariant. Let $\hat\rho$ be the density matrix on $\mathbb C^3$ that corresponds to the state $\langle \cdot \rangle^{\rm extr}_P$ restricted to the site $i=0$. We can cast the above expression as
\begin{equation}
\label{towards HCIZ}
    \int_{\mathcal U(3)} {\rm e}^{h \langle U^{-1} M_{i=0} U \rangle^{\rm extr}_P} \, {\rm d}U = {\rm e}^{-\frac h3} \int_{\mathcal U(3)} {\rm e}^{h {\rm Tr} \; U^{-1} P U \hat\rho} \, {\rm d}U.
\end{equation}
The right side depends on the eigenvalues of $hP$ and of $\hat\rho$. The eigenvalues of $hP$ are obviously $h,0,0$. In order to find those of $\hat\rho$, let us write
\begin{equation}
    \hat\rho = \sum_{a=1}^3 \rho_a |a\rangle \langle a|,
\end{equation}
where the eigenvalues satisfy $\rho_1 \geq \rho_2 \geq \rho_3 \geq 0$ and they add up to 1. Now if $|w\rangle$ is a vector in $\bbC^3$, we have
\begin{equation}
\label{expect spin}
    \bigl\langle |w\rangle \langle w| \bigr\rangle^{\rm extr}_P = \sum_{a=1}^3 \rho_a \big|\langle w | a \rangle \big|^2.
\end{equation}
Let $|v\rangle$ be the vector such that $P = |v\rangle \langle v|$.
Since $\langle\cdot\rangle^{\rm extr}_P$ is the Gibbs state that encourages the spins to orient in the direction of $|v\rangle$, the left side of \eqref{expect spin} should be largest if and only if $|w\rangle = |v\rangle$. It follows that $|1\rangle = |v\rangle$. We can relate the largest eigenvalue of $\hat\rho$ to the order parameter \eqref{def m}:
\begin{equation}
    \rho_1 = \Tr (|1\rangle \langle 1| \hat\rho) = \bigl\langle |v\rangle \langle v|_{i=0} \bigr\rangle^{\rm extr}_P = m(\beta) + \tfrac13.
\end{equation}
Next, we observe that $\langle |2\rangle \langle 2|_{i=0} \rangle^{\rm extr}_P = \langle |3\rangle \langle 3|_{i=0} \rangle^{\rm extr}_P$ by symmetry. Since $\Tr \hat\rho = 1$, we obtain all the eigenvalues of $\hat\rho$:
\begin{equation}
    \rho_1 = \tfrac13 + m(\beta); \qquad \rho_2 = \rho_3 = \tfrac13 - \tfrac12 m(\beta).
\end{equation}
The integral over $\mathcal U(3)$ can be calculated with the {\it Harish-Chandra--Itzykson--Zuber formula} \cite{Itzykson1980} -- when $P$ and $Q$ are rank-1 projectors on $\bbC^3$, the formula gives
\begin{equation}
    \int_{\mathcal U(3)} {\rm e}^{h \Tr U^{-1} P U Q} {\rm d}U = \tfrac2{h^2} ({\rm e}^h - 1 - h).
\end{equation}
We can apply this formula after subtracting a diagonal term from $\hat\rho$ in Eq.\ \eqref{towards HCIZ}. We obtain
\begin{equation}
\label{gen fct}
\begin{split}
    &\Bigl\langle \exp \Bigl\{ \frac h{|\Lambda|} \sum_{i \in \Lambda} M_i \Bigr\} \Bigr\rangle \\
    &= 2 {\rm e}^{-\frac12 hm(\beta)} \frac1{(\frac32 hm(\beta))^2} \bigl[{\rm e}^{\frac32 hm(\beta)} - 1 - \tfrac32 hm(\beta) \bigr],
\end{split}
\end{equation}
which yields the characteristic function upon setting  $h = {\rm i}k$. After taking the inverse Fourier transform, we thus obtain the order parameter distribution function
\begin{equation}
\label{order parameter fct}
    \rho(s) = \begin{cases} \frac89 \frac{m(\beta)-s}{m(\beta)^2} & \text{if } -\frac12 m(\beta) < s < m(\beta), \\ 0 & \text{otherwise.} \end{cases}
\end{equation}
We next show how this result is also obtained within the PD conjecture approach. 

\subsection{PD conjecture approach}\label{Sec:PD}

The loop model description offers useful insights regarding the quantum spin model. In particular, here it allows us to verify the symmetry breaking assumption from the previous section. PD calculations are more involved,  but their heuristics is solid and they thus yield reliable results for $\rho$. Notice that this approach can also be used within the nematic phase, see \cite{Caci2023}, where the nature of the extremal Gibbs states (for the "planar nematic" phase) is less obvious.

The PD calculations proceed from the following observations.
\begin{itemize}
    \item The quantum spin model can be viewed as a model of random loops/permutations. The average of the order parameter in the Gibbs state can be written as an expectation value of a function of loop/cycle lengths. (See Eqs. \eqref{expansion}--\eqref{cycle expression} below.)
    \item The loop model has short and long loops. The long loops scale with the volume of the system. When divided by the volume, the lengths of long loops form a random partition with a particular PD distribution. (See Eq.\ \eqref{PD conjecture} below.)
    \item We can invoke a formula for expectation values for  PD distributions, which  allows us to obtain the characteristic function of the order parameter. (See Eqs. \eqref{calc PD 1}--\eqref{result PD} below.)
\end{itemize}
The loop model is also a model of ``random transpositions". Indeed, starting from the expression \eqref{Eq:hamP} of the Hamiltonian, the partition function is equal to
\begin{equation}
\label{expansion}
\begin{split}
    &{\rm Tr} \, {\rm e}^{-\beta H} = {\rm e}^{\beta J |\mathcal B_\Lambda|} {\rm Tr} \; {\rm e}^{\beta J \sum_{\langle i,j \rangle \in \mathcal B_\Lambda} T_{i,j}} \\
    &= {\rm e}^{\beta J |\mathcal B_\Lambda|} \sum_{k\geq0} \frac{(\beta J)^k}{k!} \sum_{\langle i_1,j_1\rangle, \dots, \langle i_k,j_k \rangle} {\rm Tr} \; T_{i_1,j_1} \dots T_{i_k,j_k}.
\end{split}
\end{equation}
In order to calculate this expression, consider the permutation of the sites of $\Lambda$ that is given by the product of the transpositions of $(i_1,j_1), \dots, (i_kj_k)$, and let $\mathcal C$ denote its set of cycles. We find that
\begin{equation}
    {\rm Tr} \; T_{i_1,j_1} \dots T_{i_k,j_k} = 3^{|\mathcal C|}.
\end{equation}
Indeed, let us represent the permutation using $k$ copies of $\Lambda$ with marks to denote the transpositions. The trace amounts to summing over spin configurations where the spin values are constant in each permutation cycle. Then each cycle has three possible values of spins, independently of the values in other cycles, resulting in the number $3^{|\mathcal C|}$ above. 

Further, using again $M_i = |0\rangle \langle 0|_i- \frac13$ and working in the usual basis of spins, a similar calculation gives
\begin{equation}
\begin{split}
    {\rm Tr} \; {\rm e}^{\frac h{|\Lambda|} \sum_{i \in \Lambda} M_i} {\rm e}^{-\beta H} = & {\rm e}^{\beta J |\mathcal B_\Lambda|} \sum_{k\geq0} \frac{(\beta J)^k}{k!} \sum_{\langle i_1,j_1\rangle, \dots, \langle i_k,j_k \rangle} \\
    &\times \prod_{c \in \mathcal C} \Bigl( 2 {\rm e}^{-\frac{h}{3|\Lambda|} \ell(c)} + {\rm e}^{\frac{2h}{|\Lambda|} \ell(c)} \Bigr).
\end{split}
\end{equation}
Here, the product is over the cycles of the permutation given by the product of transpositions and $\ell(c)$ denotes the length of the cycle $c$. We have obtained
\begin{equation}
\label{cycle expression}
    \Bigl\langle \exp \Bigl\{ \frac h{|\Lambda|} \sum_{i \in \Lambda} M_i \Bigr\} \Bigr\rangle = {\mathbb E}_\Lambda \Bigl[ \prod_{c \in \mathcal C} \tfrac23 {\rm e}^{-\frac{h}{3|\Lambda|} \ell(c)} + \tfrac13 {\rm e}^{\frac{2h}{|\Lambda|} \ell(c)} \Bigr].
\end{equation}
The expectation value (average) on the right hand side is performed over permutations given by choosing arbitrary transpositions of sites in $\mathcal B_\Lambda$ (the number of transpositions is a Poisson random variable with parameter $\beta J |\mathcal B_\Lambda|$), weighted with $3^{|\mathcal C|}$. Notice that the function of cycle lengths, $f(s) = \frac23 {\rm e}^{-\frac13 s} + \frac13 {\rm e}^{\frac23 s}$, satisfies $f(0)=1$ and $f'(0)=0$; then only macroscopic cycles (whose lengths scale like $|\Lambda|$), stay relevant in the limit of large volumes. Let us introduce $\eta(\beta)$, the fraction of sites in long cycles. One expects $\eta(\beta)$ to be strictly positive when $d \geq 3$ and $\beta$ is large. The PD conjecture states that there exists $\eta(\beta) \geq 0$ such that (in the limit of infinite volume)
\begin{equation}
\label{PD conjecture}
    {\mathbb E}_\Lambda \Bigl[ \prod_{c \in \mathcal C} f\bigl( \tfrac{h}{|\Lambda|} \ell(c) \bigr) \Bigr] = {\mathbb E}_{{\rm PD}(3)} \Bigl[ \prod_{j \geq 1} f\bigl( h \eta(\beta) \lambda_j \bigr) \Bigr].
\end{equation}
The latter expectation is over partitions $(\lambda_1,\lambda_2,\dots)$ of $[0,1]$ with PD distribution. The parameter of the PD distribution is 3, a non-trivial fact that requires a separate argument to identify it. (The argument exploits the fact that $\mathbb E_\Lambda$ is invariant under a suitable Monte-Carlo evolution, and PD(3) is invariant under a suitable split-merge process; see \cite[Section II in Supplement]{Caci2023} for details).

In order to calculate the expectation of the function $f$ with respect to PD(3) we first write
\begin{equation}
\label{calc PD 1}
\begin{split}
    {\mathbb E}_{{\rm PD}(3)} &\Bigl[ \prod_{j \geq 1} f\bigl( h \eta(\beta) \lambda_j \bigr) \Bigr] \\
    &= {\rm e}^{-\frac13 h \eta(\beta)}{\mathbb E}_{{\rm PD}(3)} \Bigl[ \prod_{j \geq 1} \bigl( \tfrac23 + \tfrac13 {\rm e}^{h \eta(\beta) \lambda_j} \bigr) \Bigr].
\end{split}
\end{equation}
Now we use \cite[Eq.\ (4.16)]{Ueltschi2017}; the Taylor coefficients of the function in the expectation value are $a_\ell = \frac{(h z(\beta))^\ell}{3\ell!}$. We then get
\begin{equation}
\begin{split}
    &{\mathbb E}_{{\rm PD}(3)} \Bigl[ \prod_{j \geq 1} \bigl( \tfrac23 + \tfrac13 {\rm e}^{h \eta(\beta) \lambda_j} \bigr) \Bigr] \\
    &= 1 + 2\sum_{n\geq1} \frac1{n!} \sum_{k_1,\dots,k_n \geq 1} \frac{(h \eta(\beta))^{k_1+\dots+k_n} \Gamma(k_1) \dots \Gamma(k_n)}{k_1! \dots k_n! (k_1+\dots+k_n+2)!} \\
    &= 1 + 2\sum_{r\geq1} \frac{h \eta(\beta))^r}{(r+2)!} \underbrace{\sum_{n\geq1} \frac1{n!} \sum_{\substack{k_1,\dots,k_n\geq 1 \\ k_1+\dots+k_n=r}} \frac1{k_1 \dots k_n}}_{=1} \\
    &= 2 \sum_{r\geq0} \frac{(h \eta(\beta))^r}{(r+2)!}.
\end{split}
\end{equation}
We have thus found an expression for 
\begin{equation}
\label{result PD}
    \Bigl\langle \exp \Bigl\{ \frac h{|\Lambda|} \sum_{i \in \Lambda} M_i \Bigr\} \Bigr\rangle = 2 {\rm e}^{-\frac13 h \eta(\beta)} \sum_{r\geq0} \frac{(h \eta(\beta))^r}{(r+2)!},
\end{equation}
which yields the same result as was obtained by the symmetry breaking calculation, see Eq.\ \eqref{gen fct}, if we choose $m(\beta) = \frac23 \eta(\beta)$ (this relation could have been anticipated using calculations with permutation cycles). We finally note that the setting and the calculations generalize to arbitrary spin $S$. The transposition operator has SU($N)$ symmetry with $N=2S+1$ (the case $S=\frac12$ is the usual Heisenberg model). The PD parameter is then equal to $N$. The resulting order parameter distribution function thus depends on $N$. Here we restricted ourselves to the case of spin-1, where $N=3$.

\subsection{Quantum Monte Carlo}\label{QMC}
We next review the QMC approach that is used in our investigations. 
The stochastic series expansion (SSE) quantum Monte Carlo method with directed loop updates~\cite{Sandvik1991,Sandvik1999,Syljuasen2002, Alet2005} is an unbiased method to simulate large-scale quantum many-body systems. It
is formulated in terms of a suitable computational basis, a complete set of basis states for the Hilbert space of the system. For the spin-1 model of Eq.~(\ref{Eq:ham}), SSE simulations proceed  in the usual product basis of  local $S^z$-eigenstates. 
Observables that are diagonal with respect to this basis can then be conveniently measured. In particular, the distribution function of the order parameter $\rho(s)$ and its moments are directly accessible. For  details on the SSE method, we refer to Refs.~\cite{Sandvik1991,Sandvik1999,Syljuasen2002, Alet2005}.

At continuous phase transitions or  at strongly first-order transitions,
the characteristic properties of the phase transition can be reliably identified based on 
finite lattice simulations. The scenario of a weakly first-order phase transition is
more challenging, as the transition may appear continuous if the linear system size is
much smaller than the correlation length scale. 
Recently, a simulation protocol was
proposed to more reliably discern  such  weakly first-order phase transitions: It is based on coupling an external ordering 
field to the order parameter~\cite{Demidio2021}. In the following, we briefly
outline this approach  (cf. Ref.~\cite{Demidio2021} for details on both the method and the particular finite-size behavior): Let $H(\lambda) = H - \lambda M$ denote the Hamiltonian with an additional ordering field of
strength $\lambda$ coupled to  $M=\sum_i M_i$, and $\langle \cdot
\rangle_{\beta,\lambda}$ thermal expectation values at inverse temperature
$\beta$ with respect to $H(\lambda)$. At the critical point $\beta=\beta_c$ 
of a continuous phase transition, the order parameter vanishes as $\langle M
\rangle_{\beta_c,\lambda} \sim \lambda^{1/\delta}$ for $\lambda \to 0^+$, where
$\delta$ defines a universal critical exponent. At a first-order phase transition point, it instead 
converges towards a finite value due to phase coexistence. The
\textit{running exponent}, defined as the logarithmic
derivative
\begin{equation}\label{Eq:1oddef}
    [1/\delta](\lambda) = \frac{\partial \log \langle M \rangle_{\beta_c,\lambda}}{\partial \log \lambda} \,  ,
\end{equation}
taken at the transition temperature, thus  displays qualitatively
different behavior as $\lambda
\to 0^+$, depending on the nature of the phase transition~\cite{Demidio2021}. In the thermodynamic limit, it converges to $1/\delta$ for a continuous phase transition, and vanishes at a first-order transition point. For a finite
lattice, $[1/\delta](\lambda)$ eventually bends upward for  $\lambda \to 0^+$ in either case~\cite{Demidio2021}. However, upon increasing the system size, its
characteristic behavior of showing either a pronounced maximum prior to the upwards-bend (indicating a first-order transition) or approaching towards a plateau at a finite value prior to the upwards-bend (indicating a continuous transition) 
is still observable for sufficiently large systems~\cite{Demidio2021}. 
Namely, the running exponent may  display the characteristics of the
phase transition based on finite ordering fields $\lambda$, where the correlation length may be sufficiently reduced. Indeed, 
in Ref.~\cite{Demidio2021}, such an advancement was observed for the square lattice $J-Q$ model, establishing its weakly first-order character. 

The authors of Ref.~\cite{Demidio2021} also considered the 2D Potts model, based on calculations for its (1+1)-dimensional (1D) quantum version -- corresponding to the Hamiltonian formulation of an anisotropic classical 2D Potts model. The numerical calculations in Ref.~\cite{Demidio2021} were performed using infinite system matrix product state calculations directly in the thermodynamic limit. This leaves open the question, how  the approach proposed in  Ref.~\cite{Demidio2021}  performs, based on finite-size data for the classical 2D Potts model. In particular, one would like to know, which system sizes are required in order to  discern the rather weakly first-order character of the 5- or 6-states 2D Potts models as compared to the second order transition of  the $q$-states Potts model for $q=2,3,4$. In App.~\ref{Sec:Potts}, we report our results for the 2D Potts models from finite-size Monte Carlo simulations, in order to asses the feasibility of the approach in  Ref.~\cite{Demidio2021} to finite-size studies on this benchmark system. 
As a further benchmark example in 3D, in App.~\ref{Sec:XY}, we study the classical 3D XY model with purely biquadratic exchange interactions. For this system, the thermal phase transition is expected to belong to the 3D XY universality class~\cite{Nagata2001} (i.e., the same as the standard 3D bilinear XY model), and we find that the approach of Ref.~\cite{Demidio2021} yields results for $1/\delta$ in accord with this expectation, namely, $\delta=4.780(2)$~\cite{Campostrini2001}.  This ability of the running exponent approach to resolve this value for a 3D biquadratic model will be relevant to our study of the spin-1 system of Eq.~\ref{Eq:ham}.

Returning to  the spin-1 system of Eq.~\ref{Eq:ham}, we find that  $[1/\delta](\lambda)$
is directly accessible within the SSE. The derivation of the corresponding SSE estimator is in fact analogous to the case considered in Ref.~\cite{Demidio2021}. Within the SSE, the Hamiltonian $H$ is decomposed into a sum of bond operators, and each SSE Monte Carlo configuration contains an ordered subset of bond operator matrix elements. Denoting by 
$N^d_0$ the number of diagonal bond operator
matrix elements with at least one of the two  spins on the corresponding bond being in the local $m=0$ state (counting twice bond operator matrix elements, for which {\it both} spins are in the local $m=0$ state),
the SSE estimator of the running exponent, as obtained by the logarithmic derivative, reads
\begin{equation}\label{runningexponentspin1}
    [1/\delta](\lambda) =  \frac{\left< M N^d_0\right>_{\beta_c,\lambda}-\left< M \right>_{\beta_c,\lambda}\left< N^d_0\right>_{\beta_c,\lambda}
    }{\langle M \rangle_{\beta_c,\lambda}}   \, .
\end{equation}
We will use this direct SSE estimator in our numerical study,  i.e., without requiring any numerical differentiation.

\section{Order Parameter Distribution}\label{Sec:OP}
In Sec.~\ref{Sec:Methods}, we derived the 
order parameter distribution function of the SU(3)-symmetric spin-1 model,  
\begin{equation}
    \rho(s) = \begin{cases} \frac89 \frac{m(\beta)-s}{m(\beta)^2} & \text{if } -\frac12 m(\beta) < s < m(\beta), \\ 0 & \text{otherwise,} \end{cases}
    \label{eq:rho_S}
\end{equation}
based on  both symmetry breaking calculations and the PD conjecture approach. Here, the distribution function is expressed in terms of the expectation value $m(\beta)$ of the order parameter, which is finite in the low-$T$ ordered phase. Within the loop model description, this quantity is furthermore related as $m(\beta) = \frac23 \eta(\beta)$ to the fraction of sites contained in long loops, cf. Sec.~\ref{Sec:PD}.
From this result, we obtain the moments of the order parameter distribution function as
\begin{equation}
\begin{split}
    &\mu_k=\Bigl\langle \Bigl\{ \frac1{|\Lambda|} \sum_{i \in \Lambda} M_i \Bigr\}^k \Bigr\rangle = \int_{-\infty}^\infty s^k \rho(s) {\rm d} s \\
    &= \frac89 \frac{m(\beta)^k}{(k+1)(k+2)} \bigl[ 1 + (-\tfrac12)^{k+2} (3k+5) \bigr].    
\end{split}
\end{equation}
In particular, the zeroth-moment $\mu_0=1$ expresses the normalization of $\rho$, while the first moment $\mu_1=0$ vanishes due to the SU(3) symmetry. This is similar to the more familiar case of the $\mathbb{Z}_2$-symmetric ferromagnetic Ising model, for which in the low-$T$ ordered phase both polarization directions are equally probable, such that the $\mathbb{Z}_2$-symmetric mean value of the magnetization vanishes. Instead, it is the finite second moment of the magnetization distribution that characterizes the ordered phase of the Ising model within the $\mathbb{Z}_2$-symmetric Gibbs state. Similarly, here we obtain for the second moment  of $\rho$ a finite value  $\mu_2=\frac18 m(\beta)^2$ within the ordered low-$T$ regime (where $m(\beta)\neq 0$). Furthermore, we obtain explicitly 
$\mu_3=\frac{1}{40}  m(\beta)^3$, and $\mu_4=\frac{3}{80}  m(\beta)^4$. For the Binder ratio $R=\mu_4/\mu_2^2$ we then obtain the value $R=\frac{12}{5}$, which is in fact independent of the value of $m(\beta)$ and thus provides a generic analytical prediction  within the low-$T$ ordered regime of the SU(3) symmetric spin-1 model. Correspondingly, for the Binder cumulant $U=1-\frac13 R$ this yields the value $U=\frac15$. In the following, we will compare our analytical findings to the results of unbiased QMC simulations. 

\begin{figure}[t!]
    \centering
    \includegraphics[width=\linewidth]{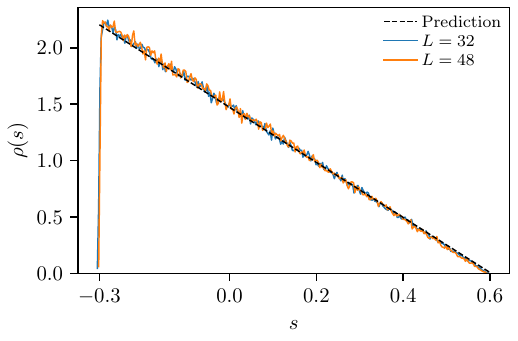}
    \caption{Order parameter distribution $\rho(s)$ obtained from QMC simulations for different linear lattice lengths $L$ at a fixed temperature $T=1/\sqrt{2}$. The black dashed line denotes the analytical prediction for $m(\beta)=0.60(2)$.}
    \label{fig:rho_s}
\end{figure}

To assess the validity of the analytical predictions, we perform QMC
simulations for different system sizes at a fixed low temperature
deep inside the ordered phase. In Fig.~\ref{fig:rho_s}, we show a comparison of
the order parameter distribution function $\rho(s)$ [cf. Eq.~(\ref{eq:rho_S})] to
the QMC simulations. Although the value of the order parameter $m(\beta)$ is
not known a priori, it can be fitted to the QMC data. We observe in 
Fig.~\ref{fig:rho_s} a remarkable good agreement of the analytical
prediction with the QMC data. Furthermore, we observe within the accuracy of our
QMC results no finite-size shifts between the $L=32$ and $L=48$ data,
indicating that $L=48$ is already representative of the thermodynamic limit for this quantity. The values of the Binder ratio $R$ that we obtain from QMC, $R=2.403(2)$ and $R=2.396(3)$ for the $L=32$ and $L=48$ system, respectively, indeed also compare well to the analytical prediction $R=12/5=2.4$. 

\section{Thermal Phase Transition}\label{Sec:TM}
In this section we examine the thermal phase transition of the SU(3) symmetric spin-1 model in Eq.~(\ref{Eq:ham}). 
In a previous QMC study \cite{Harada2002}, this transition, as well as the thermal melting of spin-nematic state in the  spin-1 model with bilinear and biquadratic terms of different magnitude, were reported to be both continuous, with similar critical exponents. However, a more recent QMC study \cite{Caci2023} showed that the spin-nematic state  melts across a weakly first-order phase transition, for which  first-order signatures emerge only for system sizes beyond those previously accessible, reminiscent of a large, but finite, correlation length. 
In the following, we  investigate the nature of the thermal melting transition of the SU(3) symmetric model by means of similarly large-scale QMC simulations.

\begin{figure}[t!]
    \centering
    \includegraphics[width=\linewidth]{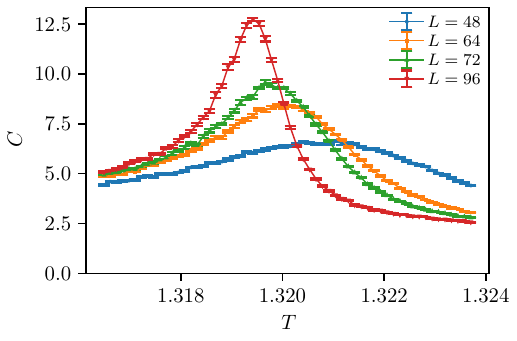}
    \caption{Specific heat $C$ as a function of temperature $T$ for different linear system sizes $L$.}
    \label{fig:C}
\end{figure}

At a first-order phase transition, the discontinuity of the energy leads to singularities in the specific heat $C$ in the thermodynamic limit. In a finite-size system, this results in a pronounced peak, for which the maximum scales extensively with the volume
$\left| \Lambda \right|$ of the system~\cite{Privman1988,Janke1993}. In Fig.~\ref{fig:C} we
show QMC results of the specific heat $C$ of the spin-1 model of  Eq.~(\ref{Eq:ham}). While we indeed observe enhanced maxima in the specific
heat upon approaching $T\approx 1.319$, these maxima, however, increase only
sub-extensively on the accessible system sizes. 

\begin{figure}[t!]
    \centering
    \includegraphics[width=\linewidth]{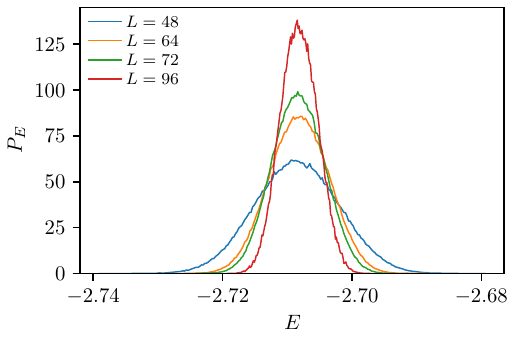}
    \caption{Energy histograms $P_E$ for different linear system sizes. The temperature is fixed for each lattice size to the value where the specific heat is maximal.}
    \label{fig:P_E}
\end{figure}

Another characteristic feature of a first-order transition is phase coexistence at the transition temperature. For example, one can analyze the energy histogram $P_E$, for which a two-peak structure is indicative of phase coexistence~\cite{Lee1990,Lee1991}. In Fig.~\ref{fig:P_E} we show QMC results of $P_E$ for different system sizes. The temperature was fixed for each finite system  to the value at which $C$ is maximal. From this data, we fail to identify any signature of phase coexistence up to the largest accessed system sizes.

\begin{figure}[t!]
    \centering
    \includegraphics[width=\linewidth]{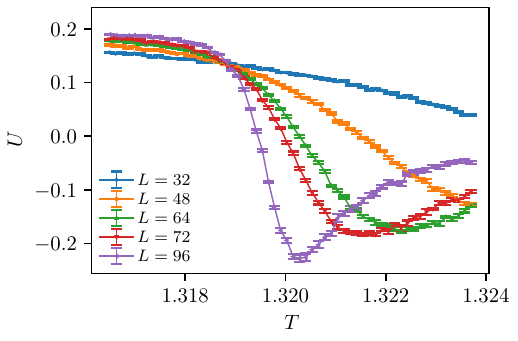}
    \caption{Temperature dependence  of the Binder cumulant $U$ for different linear system sizes $L$.}
    \label{fig:U}
\end{figure}

Next, we investigate the Binder cumulant $U$, which is shown as a function of $T$ in
Fig.~\ref{fig:U}. While the Binder cumulant shows negative dips upon
approaching a common crossing point near $T\approx 1.319$, which is usually expected for a
first-order phase transition~\cite{Janke1993}, we observe, similar to the maxima of the specific
heat, the absence of an extensive scaling of these dip values. Based on the evidence thus far, one could thus be considering a continuous phase transition near $T\approx 1.319$.

\begin{figure}[t!]
    \centering
    \includegraphics[width=\linewidth]{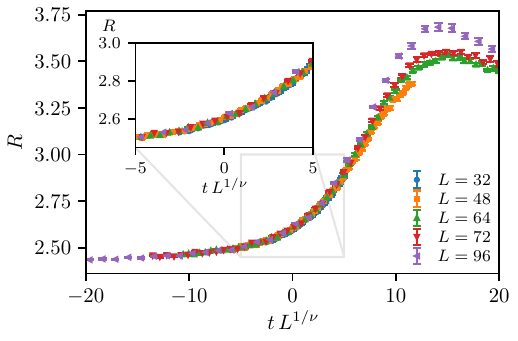}
    \caption{Data collapse of Binder ratio $R$ for different linear system sizes $L$.}
    \label{fig:R_collapse}
\end{figure}

For continuous phase transitions, a well established approach is to perform  data collapses of thermal observables near the critical point,  in order to extract the underlying critical properties. For this purpose, the finite-size data for different system sizes are rescaled 
with respect to the reduced temperature $t=(T-T_c)/T_c$, based on appropriate scaling exponents. In the following, we consider in particular the Binder ratio $R$,  which is a dimensionless quantity. Indeed, we observe in Fig.~\ref{fig:R_collapse} that the QMC data collapses,
consistent with a finite-size scaling signature  for a continuous phase
transition, for  values of  $T_c=1.3189(1)$ and $\nu=0.48(2)$ which are in accord with the QMC results of Ref.~\cite{Harada2002}. Moreover,
we  find that also other thermal observables allow for data collapses of similar quality near $T_c$, as would be expected for a continuous phase transition. Such a collapse analysis yields  estimates for the scaling exponents, namely, we obtain $\alpha=0.35(1)$ from the specific heat, $\gamma=1.00(1)$ from the order parameter  susceptibility, and $\beta=0.22(1)$ from the order parameter data collapse itself.  While the numerical values of these exponents satisfy the scaling relation $d\nu=2\beta + \gamma$ for $d=3$ rather well within the estimated uncertainty, the Rushbrooke scaling relation $ \alpha + 2\beta + \gamma= 2$ is not satisfied for  the estimated exponents. Another observation is that based on the scaling relation $2-\eta = \gamma /\nu$, we obtain a negative estimate for the anomalous exponent $\eta$, that resides close to, but below the lower bound $\eta \geq 0$ for unitary CFTs~\cite{Rychkov2017}. 
Thus, while various thermal observables show signatures of a continuous phase transition, the extracted scaling exponents are not consistent with the standard scaling relations and unitarity bounds. This is indicative of pseudoscaling,  arising near a weakly first-order phase transition due to a large correlation length, instead of genuine critical behavior. 

\begin{figure}[t!]
    \centering
    \includegraphics[width=\linewidth]{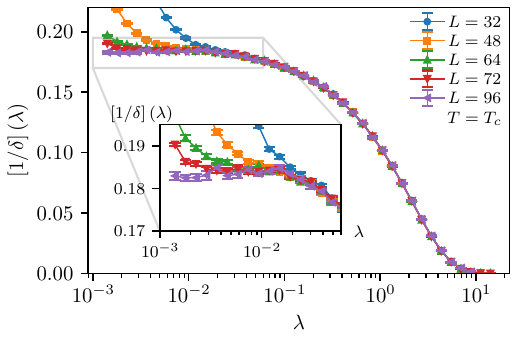}
    \caption{Flow of running exponent $[1/\delta](\lambda)$ as function of field strength $\lambda$ for different linear system sizes $L$, with  temperature  fixed to $T_c$.}
    \label{fig:invdelta_su3}
\end{figure}

The above analysis of critical exponents can be further refined by an analysis of the running exponent  $[1/\delta](\lambda)$, following Ref.~\cite{Demidio2021}, based on the direct SSE estimator in Eq.~(\ref{runningexponentspin1}). Our QMC results for this quantity are shown in Fig.~\ref{fig:invdelta_su3}. Interestingly, we find that upon increasing $L$, the QMC data appears to saturate towards a plateau value of $1/\delta = 0.184(2)$, instead of exhibiting within the accessible system sizes a  maximum that is followed by a trend towards zero for $\lambda\rightarrow 0$. However, this value of  $1/\delta$  falls significantly  below the bound $1/\delta \geq 1/5$ in $d=3$ (which relates via the scaling relation $2-\eta=d (\delta-1)/(\delta+1)$ to the bound $\eta\geq 0$ quoted above). It is important to note that the QMC results for the spin-1 model, if taken as an estimate for $1/\delta$,  violate the unitarity bound, whereas a similarly close value of $1/\delta$ to the threshold value $1/5$, but in accord with this bound, can still be resolved, e.g., for the classical biquadratic XY model considered in App.~\ref{Sec:XY}. These observations thus strengthen the versatility of the running exponent method as a valuable tool also for analyzing phase transitions. 

From our QMC simulations, we thus conclude that the thermal melting transition of  the 3D SU(3)-symmetric  spin-1 model is either weakly first-order, or, if it is continuous, it falls beyond the scope of unitary CFTs. While our results cannot fully exclude the latter scenario, in combination with our scaling analysis, the scenario of a weakly first-order phase transition appears to be the more natural scenario -- also in view of similarly weakly first-order transitions observed in the 3D CP${}^{2}$ model~\cite{Pelissetto2019}. Such a scenario furthermore appears in line with the observations in Ref.~\cite{Caci2023}, according to which 
the spin-nematic thermal melting transition is weakly first-order as well. In fact, as shown in App.~\ref{Sec:Nematic}, the running exponent approach confirms the first-order character of the spin-nematic thermal melting transition of the 3D spin-1 model. 

\section{Conclusions}
We examined both the ordered phase and the thermal phase transition of the 3D spin-1
Heisenberg model on the simple cubic lattice with bilinear and biquadratic interactions of equal strength, giving rise to an enhanced SU(3) symmetric Hamiltonian. Based on both Poisson-Dirichlet and symmetry-breaking calculations, we derived analytically the order parameter distribution function and compared these analytic results to unbiased quantum Monte Carlo simulations. For both the distribution function itself and the corresponding Binder cumulant ratio, we observe good agreement with the numerical data. Furthermore, we used quantum Monte Carlo simulations, combined with finite-size analysis,  in order to determine the nature of the thermal melting transition, which was previously claimed to be continuous~\cite{Harada2002}. While we do not obtain  direct indications from conventional probes for a first-order phase transition, such as in the finite-size scaling of the specific heat maximum or a two-peak structure in energy histograms, we obtain an indirect indication from using the running exponent method of Ref.~\cite{Demidio2021}: In case of a continuous transition, our numerical data would point towards a value of the critical exponent $1/\delta$ that falls below the lower bound of $1/5$ for unitary conformal critical points in 3D. Instead, a scenario, in which the thermal melting is in fact weakly first order, but with a rather large correlation length, i.e., beyond the scale of the  system sizes accessible to us, appears to be the more natural explanation of our findings. This would imply that for even larger system sizes than those accessed here, the apparent plateau value in the running exponent $[1/\delta](\lambda)$ would be resolved as a very broad maximum, similar to the behavior that we observe for the 2D 5-states Potts model (which is known to exhibit a weakly first-order transition). For the future, it would thus be valuable to better understand under which conditions, such as, e.g., anisotropic couplings, one can more efficiently reduce the relevant correlation length scales at 2D and 3D first-order phase transitions in order to identify their nature, similar to what is apparently the case for the 2D $J-Q$ model and the 1D quantum Potts model~\cite{Demidio2021}.

We conclude with a brief discussion of another spin-1 model that has nearest-neighbor interactions with an enhanced SU(3) symmetry (only) on bipartite graphs (cf. also the discussion in Ref.~\cite{Ueltschi2015}). The Hamiltonian is given in terms of a purely biquadratic exchange interaction: 
\begin{equation}
\check{\Ham} = -\: J \!\!\sum_{\langle i,j\rangle \in\caB_\Lambda} (\mathbf{S}_i \cdot \mathbf{S}_j)^2.
\end{equation}
This interaction term is related to the projector $P_{i,j}^{(0)}$ onto the one-dimensional subspace where $(\mathbf{S}_i + \mathbf{S}_j)^2$ has eigenvalue 0, namely, $(\mathbf{S}_i \cdot \mathbf{S}_j)^2 = 3P_{i,j}^{(0)} + 1$. In the local $S^z$-basis, 
\begin{equation}
    P_{i,j}^{(0)} = \tfrac13 \sum_{a,b=-1}^1 (-1)^{a+b} |a,-a\rangle \langle b,-b|.
\end{equation}
Locally, the projected state is maximally entangled and the ground state of $\check H$ on the full lattice is therefore rather complicated. We expect nonetheless that for $J>0$ this model behaves similarly to the model described in this paper. Namely, the SU(3) symmetry is broken at low temperatures and the extremal Gibbs states are labeled by rank-1 projectors. (The definition is similar to Eq.\ \eqref{extremal states} but the external field has a staggered nature.) The order parameter can be chosen to be similar to $m(\beta)$ in Eq.\ \eqref{def m}, taking into account that the Gibbs states are periodic rather than translation-invariant. 
The symmetry breaking calculations, and Eqs. \eqref{gen fct} and \eqref{order parameter fct} remain valid. The loop representation is different and was proposed by Aizenman and Nachtergaele \cite{Aizenman1994}, but the behavior is similar. The Poisson-Dirichlet conjecture still holds with parameter 3 and Eqs. \eqref{cycle expression}--\eqref{result PD} remain valid. 
The thermal phase transition of $\check{\Ham}$ on the simple cubic lattice was located
in Ref.~\cite{Harada2002}
at a temperature of $T=1.5652(7)J$, and concluded to be continuous. 
In view of our findings,  it would certainly be interesting to re-examine also the case of $\check{\Ham}$ in more detail in future studies. 

Let us finally remark that the case $J<0$, for $H$ and $\check H$, are also interesting but very difficult to study: The loop representations carry signs and the probabilistic nature is lost; the QMC method suffers from a severe sign problem.

\begin{acknowledgments}
We thank J. D’Emidio, A. L\"auchli, and  F. Parisen Toldin for discussions. Furthermore, we acknowledge computing time granted by the IT Center of RWTH Aachen University. NC acknowledges support by the ANR through grant LODIS (ANR-21-CE30-0033).
\end{acknowledgments}

\appendix 

\section{2D Potts model}\label{Sec:Potts}

In this appendix, we examine the application of  the running exponent approach of Ref.~\cite{Demidio2021} to the 2D Potts model. For our finite-size Monte Carlo study,  we consider  square lattices with $N=L^2$ sites, and local discrete variables $\sigma_i=1,\dots,q$, in terms of which the $q$-states Potts model is given by
\begin{equation}
    H_P=-J\sum_{\langle i,j \rangle} \delta_{\sigma_i,\sigma_j}-\lambda\sum_i \delta_{\sigma_i,1}\,.
\end{equation}
Here, $\langle i,j \rangle$ denotes nearest neighbor bonds and $\delta_{\sigma_i,\sigma_j}$ is the Kronecker delta, and we consider periodic boundary conditions.  In addition to the interaction term, an external field of strength $\lambda$ is introduced in order to apply the approach of Ref.~\cite{Demidio2021}, and which favors the state $\sigma_i=1$ on each lattice site. In the absence of the external field, i.e., for $\lambda=0$, the 2D Potts model is well known to exhibit a thermal ordering transition in the thermodynamic limit, which is continuous for $q\le 4$, and first-order for $q>4$~\cite{Wu1982}. In either case, the critical temperature is given by
\begin{equation}
T_c=\frac{J}{\log(1+\sqrt{q})}\,.
\end{equation}

\begin{figure*}[t!]
    \centering
    \includegraphics[width=\textwidth]{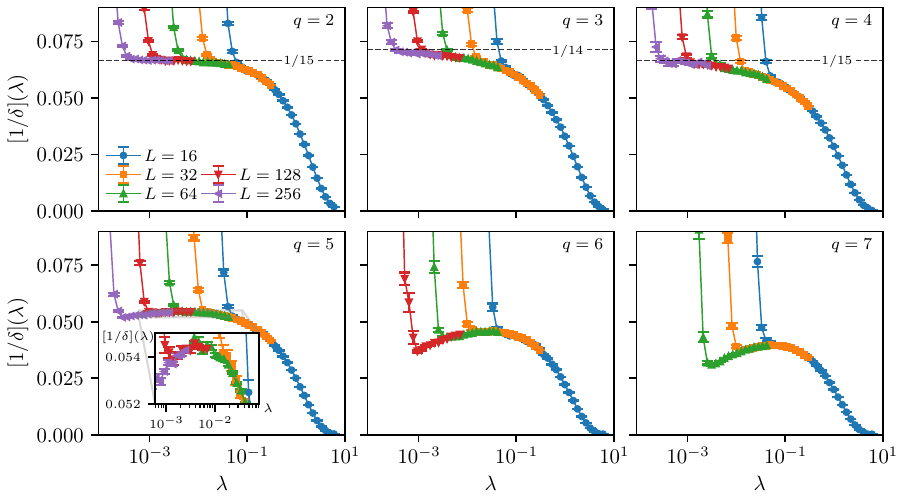}
    \caption{Running exponent $[1/\delta](\lambda)$ as a function of $\lambda$ for various 2D $q$-states Potts models for different linear system sizes $L$ at the respective transition temperature.}
    \label{fig:potts}
\end{figure*}

In the following, we fix $J=1$. Since a finite external field $\lambda >0$ favors the local state $\sigma_i=1$, we consider the order parameter
\begin{equation}
M=\frac{q n_1-1}{q-1},
\end{equation}
in terms of the density $n_1$ of the local $\sigma_i=1$ (i.e., on a finite lattice, $n_1=\langle N_1/N\rangle_{\beta,\lambda}$ is given by the expectation value of the number $N_1$ of lattice sites in the local state $\sigma_i=1$). While $M\rightarrow0$ for $T\rightarrow \infty$, at low temperatures, $M\rightarrow1$ for $T\rightarrow 0$ and $\lambda>0$. For the running exponent $[1/\delta](\lambda)$, defined as in Eq.~(\ref{Eq:1oddef}), we obtain the explicit estimator
\begin{equation}
    [1/\delta](\lambda)=\lambda \beta_c N  \frac{\langle n_1 [qn_1 \! - \! 1]\rangle_{\beta_c,\lambda} - \langle n_1\rangle_{\beta_c,\lambda} \langle qn_1 \! - \! 1\rangle_{\beta_c,\lambda} }{\langle q n_1\! - \! 1 \rangle_{\beta_c,\lambda}},
\end{equation}
where $\beta_c=1/T_c$. As a good quantitative consistency check, we can compare to the known values of the critical exponent $1/\delta$, which read $1/\delta=1/15, 1/14, 1/15$ for $q=2,3,4$, respectively~\cite{Wu1982}. In terms of the correlation length $\xi$, which has been previously estimated as $\xi=2512.2, 158.9, 48.1, 23.9$ for $q=5,6,7,8$, one finds that in particular the $q=5$ and  $q=6$ 2D Potts model  exhibit  rather weakly first-order phase transitions, with large correlation lengths~\cite{Buddenoir1993,Iino2019}. In order to perform Monte Carlo simulations of the Potts model in the presence of a finite external field, we used the Wolff cluster updates in the ghost site formulation of Ref.~\cite{Kent-Dobias2018}.

In Fig.~\ref{fig:potts}, we present our Monte Carlo results for the running exponents for various values of $q$. For $q=2,3,4$, the simulation results are in good agreement with the expectation of a continuous transition in all three cases, and  the known values of $1/\delta$ are  reproduced by the running exponent  approach in the plateau regime. Turning  to the first-order cases, $q>4$, we find that for both $q=6$ and $q=7$, we can resolve a characteristic maximum in  $[1/\delta](\lambda)$, indicating a first-order transition, on system sizes with a linear extend $L$ of order of half  the correlation length $\xi$, but not for significantly smaller lattices. For the case $q=5$, we  obtain indications for a  maximum from simulations performed for $L=128$, and up to the largest accessible system sizes $L=256$ (see also the discussion in Ref.~\cite{Iino2019} regarding system size limitations). In all cases, one thus requires to simulate finite systems of a size of similar order of magnitude as when using alternative data analysis approaches in order to ascertain the nature of the phase transition for this benchmark system~\cite{Iino2019}.

\section{3D biquadratic XY model}\label{Sec:XY}

In this appendix, we apply the running exponent approach from Ref.~\cite{Demidio2021} to the 3D biquadratic XY model on the simple cubic lattice $\Lambda$, with $N=|\Lambda|=L^3$ sites,  defined by the Hamiltonian 
\begin{equation}
H_{XY}=- J \sum_{\langle i,j \rangle} (\vec{S}_i\cdot \vec{S}_j)^2\, -\lambda\sum_i Q^{xy}_i,
\end{equation}
where $\vec{S}_i=(S_i^x,S_i^y)$ denotes a classical  real two-component  unit vector on lattice site $i$, and $Q^{xy}_i=(S^x_i)^2-(S^y_i)^2$. 
Note that besides its global O(2) symmetry for $\lambda=0$, the model also exhibits a local $\mathbb{Z}_2$ symmetry, as  $H_{XY}$ is invariant under independent local transformations $\vec{S}_i \rightarrow  -\vec{S}_i$, in contrast to the usual bilinear XY model. In the following, we will fix the interaction strength to $J=1$. We also included in $H_{XY}$ an external field of strength $\lambda$ that couples to the local nematic order parameter $Q^{xy}_i$~\cite{Nagata2001}. Indeed, it is known from previous studies, that in the absence of the external field, i.e., for $\lambda=0$, the model exhibits a thermal phase transition from a high-temperature paramagnetic phase to a 
low-temperature phase with long-range spin nematic (quadrupolar) order~\cite{Nagata2001}. Furthermore, finite-size estimates of the critical exponents $\nu$ and $\gamma$ were found in agreement with the 3D XY universality class~\cite{Nagata2001}. 

\begin{figure}[t!]
    \centering
    \includegraphics[width=\linewidth]{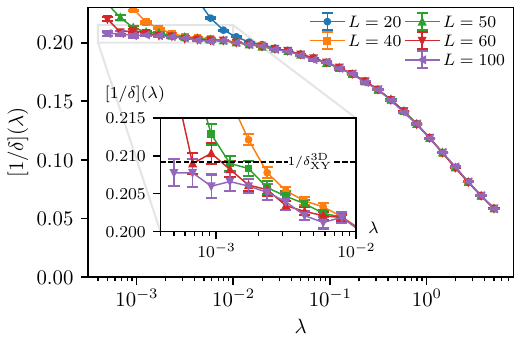}
    \caption{Running exponent $[1/\delta](\lambda)$ as a function of $\lambda$ for different linear system sizes $L$ for the critical 3D biquadratic XY model. The dashed line in the inset indicates the value of $1/\delta^\text{3D}_{XY}$, with $\delta^\text{3D}_{XY}=4.780(2)$ for the anticipated 3D XY universality class~\cite{Campostrini2001}. }
    \label{fig:xy3d}
\end{figure}

Here, we apply the running exponent method from Ref.~\cite{Demidio2021} to this model, in order to check, if it is feasible, based on this method, to obtain an independent estimate for $1/\delta$ that is in accord with the value $\delta=\delta^\text{3D}_{XY}=4.780(2)$ for the anticipated 3D XY universality class~\cite{Campostrini2001}. For this purpose, we performed finite-size simulations for different linear system sizes $L$ at a temperature of $T=1.101$, based on the estimate for the transition temperature in Ref.~\cite{Nagata2001}. From taking the logarithmic derivative of $\langle Q^{xy}\rangle_{\beta,\lambda}$ with respect to the $\lambda$, as in Eq.~(\ref{Eq:1oddef}), where $Q^{xy}=\frac{1}{N}\sum_iQ^{xy}_i$, we obtain an explicit estimator 
\begin{equation}
    [1/\delta](\lambda)=\beta_c\lambda\frac{\langle (Q^{xy})^2\rangle - \langle Q^{xy}\rangle^2}{\langle Q^{xy}\rangle}\,
\end{equation}
for the running exponent. 

Our numerical data, summarized in Fig.~\ref{fig:xy3d} indeed exhibits the expected behavior that for sufficiently large system sizes the plateau value of  $[1/\delta](\lambda)$ approaches towards the anticipated value for the 3D XY universality class, indicated by the dashed line. 

\section{3D bilinear-biquadratic spin-1 model}\label{Sec:Nematic}

In this appendix we consider the 3D bilinear-biquadratic spin-1 model on the simple cubic lattice in a regime where  the bilinear and biquadratic terms have different coupling strengths. The Hamiltonian of this model is denoted by
\begin{equation}
\label{Eq:ham_SN}
\tilde{\Ham} = -\: J \!\!\sum_{\langle i,j\rangle \in\caB_\Lambda}\!\!\left[   \, u ( \mathbf{S}_i \cdot \mathbf{S}_j )+   \, (\mathbf{S}_i \cdot \mathbf{S}_j)^2\: \right] \, ,
\end{equation}
where for $u=1$, $\tilde{H}=H$ is recovered. This model is known to feature a spin-nematic low-temperature phase in the parameter regime $u\in (0,1)$. In Ref.~\cite{Caci2023}, this spin-nematic state was found to melt across a weakly first-order phase transition.

\begin{figure}[t!]
    \centering
    \includegraphics[width=\linewidth]{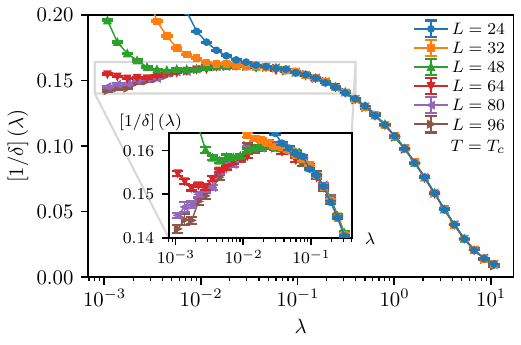}
    \caption{Running exponent $[1/\delta](\lambda)$ as a function of $\lambda$ for $u=\mathrm{cot}(3\pi/8)$ and different linear system sizes $L$. The temperature is fixed to the transition temperature $T_c$.}
    \label{fig:invdelta_nem}
\end{figure}

Here, we investigate this spin-nematic melting transition using the running exponent approach~\cite{Demidio2021}, fixing the coupling parameter to $u=\mathrm{cot}(3\pi/8)$, a value that was also considered in Ref.~\cite{Caci2023} -- chosen in view of  the more conventional trigonometric parametrization of $\tilde{H}$ (cf. Ref.~\cite{Caci2023}). In Ref.~\cite{Caci2023}, clear first-order signatures such as phase coexistence were observable for $L\gtrsim 64$ systems at this parameter value.

The order parameter of the spin-nematic state is given by $Q = \left|\Lambda\right|^{-1} \sum_i Q_i$, where $Q_i = (S_i^z)^2 - 2/3$. Within the running exponent analysis we couple this order parameter to an ordering field of strength $\lambda$, 
\begin{equation}
    \tilde{\Ham}(\lambda) =  \tilde{\Ham} + \lambda Q \, .
    \label{eq:H_lambda_SN}
\end{equation}
Here, the $+$ sign has to be chosen, due to the quantum nature of the spin-nematic state (cf. the discussion provided in Ref.~\cite{Ueltschi2015} for details). The running exponent as a function of $\lambda$ is shown in Fig.~\ref{fig:invdelta_nem} for different system sizes, and at the transition temperature $T_c=1.649$~\cite{Caci2023}. The behavior displays a pronounced maximum, indicative of a first-order phase transition. This signature is already observable for $L=48$, thus slightly lowering the lattice sizes required to identify the nature of the phase transition. 

\bibliography{paper.bbl}

\begin{thebibliography}{44}%
\makeatletter
\providecommand \@ifxundefined [1]{%
 \@ifx{#1\undefined}
}%
\providecommand \@ifnum [1]{%
 \ifnum #1\expandafter \@firstoftwo
 \else \expandafter \@secondoftwo
 \fi
}%
\providecommand \@ifx [1]{%
 \ifx #1\expandafter \@firstoftwo
 \else \expandafter \@secondoftwo
 \fi
}%
\providecommand \natexlab [1]{#1}%
\providecommand \enquote  [1]{``#1''}%
\providecommand \bibnamefont  [1]{#1}%
\providecommand \bibfnamefont [1]{#1}%
\providecommand \citenamefont [1]{#1}%
\providecommand \href@noop [0]{\@secondoftwo}%
\providecommand \href [0]{\begingroup \@sanitize@url \@href}%
\providecommand \@href[1]{\@@startlink{#1}\@@href}%
\providecommand \@@href[1]{\endgroup#1\@@endlink}%
\providecommand \@sanitize@url [0]{\catcode `\\12\catcode `\$12\catcode
  `\&12\catcode `\#12\catcode `\^12\catcode `\_12\catcode `\%12\relax}%
\providecommand \@@startlink[1]{}%
\providecommand \@@endlink[0]{}%
\providecommand \url  [0]{\begingroup\@sanitize@url \@url }%
\providecommand \@url [1]{\endgroup\@href {#1}{\urlprefix }}%
\providecommand \urlprefix  [0]{URL }%
\providecommand \Eprint [0]{\href }%
\providecommand \doibase [0]{http://dx.doi.org/}%
\providecommand \selectlanguage [0]{\@gobble}%
\providecommand \bibinfo  [0]{\@secondoftwo}%
\providecommand \bibfield  [0]{\@secondoftwo}%
\providecommand \translation [1]{[#1]}%
\providecommand \BibitemOpen [0]{}%
\providecommand \bibitemStop [0]{}%
\providecommand \bibitemNoStop [0]{.\EOS\space}%
\providecommand \EOS [0]{\spacefactor3000\relax}%
\providecommand \BibitemShut  [1]{\csname bibitem#1\endcsname}%
\let\auto@bib@innerbib\@empty
\bibitem [{\citenamefont {Kaplan}\ \emph {et~al.}(2009)\citenamefont {Kaplan},
  \citenamefont {Lee}, \citenamefont {Son},\ and\ \citenamefont
  {Stephanov}}]{Kaplan2009}%
  \BibitemOpen
  \bibfield  {author} {\bibinfo {author} {\bibfnamefont {David~B.}\
  \bibnamefont {Kaplan}}, \bibinfo {author} {\bibfnamefont {Jong-Wan}\
  \bibnamefont {Lee}}, \bibinfo {author} {\bibfnamefont {Dam~T.}\ \bibnamefont
  {Son}}, \ and\ \bibinfo {author} {\bibfnamefont {Mikhail~A.}\ \bibnamefont
  {Stephanov}},\ }\bibfield  {title} {\enquote {\bibinfo {title} {Conformality
  lost},}\ }\href {\doibase 10.1103/PhysRevD.80.125005} {\bibfield  {journal}
  {\bibinfo  {journal} {Phys. Rev. D}\ }\textbf {\bibinfo {volume} {80}},\
  \bibinfo {pages} {125005} (\bibinfo {year} {2009})}\BibitemShut {NoStop}%
\bibitem [{\citenamefont {Nahum}\ \emph {et~al.}(2015)\citenamefont {Nahum},
  \citenamefont {Chalker}, \citenamefont {Serna}, \citenamefont {Ortu\~no},\
  and\ \citenamefont {Somoza}}]{Nahum2015}%
  \BibitemOpen
  \bibfield  {author} {\bibinfo {author} {\bibfnamefont {Adam}\ \bibnamefont
  {Nahum}}, \bibinfo {author} {\bibfnamefont {J.~T.}\ \bibnamefont {Chalker}},
  \bibinfo {author} {\bibfnamefont {P.}~\bibnamefont {Serna}}, \bibinfo
  {author} {\bibfnamefont {M.}~\bibnamefont {Ortu\~no}}, \ and\ \bibinfo
  {author} {\bibfnamefont {A.~M.}\ \bibnamefont {Somoza}},\ }\bibfield  {title}
  {\enquote {\bibinfo {title} {Deconfined quantum criticality, scaling
  violations, and classical loop models},}\ }\href {\doibase
  10.1103/PhysRevX.5.041048} {\bibfield  {journal} {\bibinfo  {journal} {Phys.
  Rev. X}\ }\textbf {\bibinfo {volume} {5}},\ \bibinfo {pages} {041048}
  (\bibinfo {year} {2015})}\BibitemShut {NoStop}%
\bibitem [{\citenamefont {Wang}\ \emph {et~al.}(2017)\citenamefont {Wang},
  \citenamefont {Nahum}, \citenamefont {Metlitski}, \citenamefont {Xu},\ and\
  \citenamefont {Senthil}}]{Wang2017}%
  \BibitemOpen
  \bibfield  {author} {\bibinfo {author} {\bibfnamefont {Chong}\ \bibnamefont
  {Wang}}, \bibinfo {author} {\bibfnamefont {Adam}\ \bibnamefont {Nahum}},
  \bibinfo {author} {\bibfnamefont {Max~A.}\ \bibnamefont {Metlitski}},
  \bibinfo {author} {\bibfnamefont {Cenke}\ \bibnamefont {Xu}}, \ and\ \bibinfo
  {author} {\bibfnamefont {T.}~\bibnamefont {Senthil}},\ }\bibfield  {title}
  {\enquote {\bibinfo {title} {Deconfined quantum critical points: Symmetries
  and dualities},}\ }\href {\doibase 10.1103/PhysRevX.7.031051} {\bibfield
  {journal} {\bibinfo  {journal} {Phys. Rev. X}\ }\textbf {\bibinfo {volume}
  {7}},\ \bibinfo {pages} {031051} (\bibinfo {year} {2017})}\BibitemShut
  {NoStop}%
\bibitem [{\citenamefont {Gorbenko}\ \emph
  {et~al.}(2018{\natexlab{a}})\citenamefont {Gorbenko}, \citenamefont
  {Rychkov},\ and\ \citenamefont {Zan}}]{Gorbenko2018}%
  \BibitemOpen
  \bibfield  {author} {\bibinfo {author} {\bibfnamefont {Victor}\ \bibnamefont
  {Gorbenko}}, \bibinfo {author} {\bibfnamefont {Slava}\ \bibnamefont
  {Rychkov}}, \ and\ \bibinfo {author} {\bibfnamefont {Bernardo}\ \bibnamefont
  {Zan}},\ }\bibfield  {title} {\enquote {\bibinfo {title} {Walking, weak
  first-order transitions, and complex cfts},}\ }\href {\doibase
  10.1007/JHEP10(2018)108} {\bibfield  {journal} {\bibinfo  {journal} {Journal
  of High Energy Physics}\ }\textbf {\bibinfo {volume} {2018}},\ \bibinfo
  {pages} {108} (\bibinfo {year} {2018}{\natexlab{a}})}\BibitemShut {NoStop}%
\bibitem [{\citenamefont {Gorbenko}\ \emph
  {et~al.}(2018{\natexlab{b}})\citenamefont {Gorbenko}, \citenamefont
  {Rychkov},\ and\ \citenamefont {Zan}}]{Gorbenko2018II}%
  \BibitemOpen
  \bibfield  {author} {\bibinfo {author} {\bibfnamefont {Victor}\ \bibnamefont
  {Gorbenko}}, \bibinfo {author} {\bibfnamefont {Slava}\ \bibnamefont
  {Rychkov}}, \ and\ \bibinfo {author} {\bibfnamefont {Bernardo}\ \bibnamefont
  {Zan}},\ }\bibfield  {title} {\enquote {\bibinfo {title} {{Walking, Weak
  first-order transitions, and Complex CFTs II. Two-dimensional Potts model at
  $Q>4$}},}\ }\href {\doibase 10.21468/SciPostPhys.5.5.050} {\bibfield
  {journal} {\bibinfo  {journal} {SciPost Phys.}\ }\textbf {\bibinfo {volume}
  {5}},\ \bibinfo {pages} {50} (\bibinfo {year}
  {2018}{\natexlab{b}})}\BibitemShut {NoStop}%
\bibitem [{\citenamefont {Ma}\ and\ \citenamefont {He}(2019)}]{Ma2019}%
  \BibitemOpen
  \bibfield  {author} {\bibinfo {author} {\bibfnamefont {Han}\ \bibnamefont
  {Ma}}\ and\ \bibinfo {author} {\bibfnamefont {Yin-Chen}\ \bibnamefont {He}},\
  }\bibfield  {title} {\enquote {\bibinfo {title} {Shadow of complex fixed
  point: Approximate conformality of $q\geq 4$ potts model},}\ }\href {\doibase
  10.1103/PhysRevB.99.195130} {\bibfield  {journal} {\bibinfo  {journal} {Phys.
  Rev. B}\ }\textbf {\bibinfo {volume} {99}},\ \bibinfo {pages} {195130}
  (\bibinfo {year} {2019})}\BibitemShut {NoStop}%
\bibitem [{\citenamefont {Nogueira}\ \emph {et~al.}(2019)\citenamefont
  {Nogueira}, \citenamefont {van~den Brink},\ and\ \citenamefont
  {Sudb\o{}}}]{Nogueira2019}%
  \BibitemOpen
  \bibfield  {author} {\bibinfo {author} {\bibfnamefont {Flavio~S.}\
  \bibnamefont {Nogueira}}, \bibinfo {author} {\bibfnamefont {Jeroen}\
  \bibnamefont {van~den Brink}}, \ and\ \bibinfo {author} {\bibfnamefont
  {Asle}\ \bibnamefont {Sudb\o{}}},\ }\bibfield  {title} {\enquote {\bibinfo
  {title} {Conformality loss and quantum criticality in topological higgs
  electrodynamics in $2+1$ dimensions},}\ }\href {\doibase
  10.1103/PhysRevD.100.085005} {\bibfield  {journal} {\bibinfo  {journal}
  {Phys. Rev. D}\ }\textbf {\bibinfo {volume} {100}},\ \bibinfo {pages}
  {085005} (\bibinfo {year} {2019})}\BibitemShut {NoStop}%
\bibitem [{\citenamefont {Iino}\ \emph {et~al.}(2019)\citenamefont {Iino},
  \citenamefont {Morita}, \citenamefont {Kawashima},\ and\ \citenamefont
  {Sandvik}}]{Iino2019}%
  \BibitemOpen
  \bibfield  {author} {\bibinfo {author} {\bibfnamefont {Shumpei}\ \bibnamefont
  {Iino}}, \bibinfo {author} {\bibfnamefont {Satoshi}\ \bibnamefont {Morita}},
  \bibinfo {author} {\bibfnamefont {Naoki}\ \bibnamefont {Kawashima}}, \ and\
  \bibinfo {author} {\bibfnamefont {Anders~W.}\ \bibnamefont {Sandvik}},\
  }\bibfield  {title} {\enquote {\bibinfo {title} {Detecting signals of weakly
  first-order phase transitions in two-dimensional potts models},}\ }\href
  {\doibase 10.7566/JPSJ.88.034006} {\bibfield  {journal} {\bibinfo  {journal}
  {Journal of the Physical Society of Japan}\ }\textbf {\bibinfo {volume}
  {88}},\ \bibinfo {pages} {034006} (\bibinfo {year} {2019})}\BibitemShut
  {NoStop}%
\bibitem [{\citenamefont {D'Emidio}\ \emph {et~al.}(2023)\citenamefont
  {D'Emidio}, \citenamefont {Eberharter},\ and\ \citenamefont
  {Läuchli}}]{Demidio2021}%
  \BibitemOpen
  \bibfield  {author} {\bibinfo {author} {\bibfnamefont {Jonathan}\
  \bibnamefont {D'Emidio}}, \bibinfo {author} {\bibfnamefont {Alexander~A.}\
  \bibnamefont {Eberharter}}, \ and\ \bibinfo {author} {\bibfnamefont
  {Andreas~M.}\ \bibnamefont {Läuchli}},\ }\bibfield  {title} {\enquote
  {\bibinfo {title} {{Diagnosing weakly first-order phase transitions by
  coupling to order parameters}},}\ }\href {\doibase
  10.21468/SciPostPhys.15.2.061} {\bibfield  {journal} {\bibinfo  {journal}
  {SciPost Phys.}\ }\textbf {\bibinfo {volume} {15}},\ \bibinfo {pages} {061}
  (\bibinfo {year} {2023})}\BibitemShut {NoStop}%
\bibitem [{\citenamefont {Caci}\ \emph {et~al.}(2023)\citenamefont {Caci},
  \citenamefont {M\"uhlbacher}, \citenamefont {Ueltschi},\ and\ \citenamefont
  {Wessel}}]{Caci2023}%
  \BibitemOpen
  \bibfield  {author} {\bibinfo {author} {\bibfnamefont {Nils}\ \bibnamefont
  {Caci}}, \bibinfo {author} {\bibfnamefont {Peter}\ \bibnamefont
  {M\"uhlbacher}}, \bibinfo {author} {\bibfnamefont {Daniel}\ \bibnamefont
  {Ueltschi}}, \ and\ \bibinfo {author} {\bibfnamefont {Stefan}\ \bibnamefont
  {Wessel}},\ }\bibfield  {title} {\enquote {\bibinfo {title}
  {Poisson-dirichlet distributions and weakly first-order spin-nematic phase
  transitions},}\ }\href {\doibase 10.1103/PhysRevB.107.L020409} {\bibfield
  {journal} {\bibinfo  {journal} {Phys. Rev. B}\ }\textbf {\bibinfo {volume}
  {107}},\ \bibinfo {pages} {L020409} (\bibinfo {year} {2023})}\BibitemShut
  {NoStop}%
\bibitem [{\citenamefont {Tanaka}\ \emph {et~al.}(2001)\citenamefont {Tanaka},
  \citenamefont {Tanaka},\ and\ \citenamefont {Idogaki}}]{Tanaka_2001}%
  \BibitemOpen
  \bibfield  {author} {\bibinfo {author} {\bibfnamefont {Kengo}\ \bibnamefont
  {Tanaka}}, \bibinfo {author} {\bibfnamefont {Akinori}\ \bibnamefont
  {Tanaka}}, \ and\ \bibinfo {author} {\bibfnamefont {Toshihiro}\ \bibnamefont
  {Idogaki}},\ }\bibfield  {title} {\enquote {\bibinfo {title} {Long-range
  order in the ground state of the s = 1 isotropic bilinear-biquadratic
  exchange hamiltonian},}\ }\href {\doibase 10.1088/0305-4470/34/42/304}
  {\bibfield  {journal} {\bibinfo  {journal} {Journal of Physics A:
  Mathematical and General}\ }\textbf {\bibinfo {volume} {34}},\ \bibinfo
  {pages} {8767} (\bibinfo {year} {2001})}\BibitemShut {NoStop}%
\bibitem [{\citenamefont {Harada}\ and\ \citenamefont
  {Kawashima}(2002)}]{Harada2002}%
  \BibitemOpen
  \bibfield  {author} {\bibinfo {author} {\bibfnamefont {Kenji}\ \bibnamefont
  {Harada}}\ and\ \bibinfo {author} {\bibfnamefont {Naoki}\ \bibnamefont
  {Kawashima}},\ }\bibfield  {title} {\enquote {\bibinfo {title} {Quadrupolar
  order in isotropic heisenberg models with biquadratic interaction},}\ }\href
  {\doibase 10.1103/PhysRevB.65.052403} {\bibfield  {journal} {\bibinfo
  {journal} {Phys. Rev. B}\ }\textbf {\bibinfo {volume} {65}},\ \bibinfo
  {pages} {052403} (\bibinfo {year} {2002})}\BibitemShut {NoStop}%
\bibitem [{\citenamefont {Batista}\ and\ \citenamefont
  {Ortiz}(2004)}]{Batista2004}%
  \BibitemOpen
  \bibfield  {author} {\bibinfo {author} {\bibfnamefont {C.~D.}\ \bibnamefont
  {Batista}}\ and\ \bibinfo {author} {\bibfnamefont {G.}~\bibnamefont
  {Ortiz}},\ }\bibfield  {title} {\enquote {\bibinfo {title} {Algebraic
  approach to interacting quantum systems},}\ }\href {\doibase
  10.1080/00018730310001642086} {\bibfield  {journal} {\bibinfo  {journal}
  {Advances in Physics}\ }\textbf {\bibinfo {volume} {53}},\ \bibinfo {pages}
  {1--82} (\bibinfo {year} {2004})}\BibitemShut {NoStop}%
\bibitem [{\citenamefont {T\'oth}\ \emph {et~al.}(2012)\citenamefont {T\'oth},
  \citenamefont {L\"auchli}, \citenamefont {Mila},\ and\ \citenamefont
  {Penc}}]{Toth2012}%
  \BibitemOpen
  \bibfield  {author} {\bibinfo {author} {\bibfnamefont {Tam\'as~A.}\
  \bibnamefont {T\'oth}}, \bibinfo {author} {\bibfnamefont {Andreas~M.}\
  \bibnamefont {L\"auchli}}, \bibinfo {author} {\bibfnamefont {Fr\'ed\'eric}\
  \bibnamefont {Mila}}, \ and\ \bibinfo {author} {\bibfnamefont {Karlo}\
  \bibnamefont {Penc}},\ }\bibfield  {title} {\enquote {\bibinfo {title}
  {Competition between two- and three-sublattice ordering for $s=1$ spins on
  the square lattice},}\ }\href {\doibase 10.1103/PhysRevB.85.140403}
  {\bibfield  {journal} {\bibinfo  {journal} {Phys. Rev. B}\ }\textbf {\bibinfo
  {volume} {85}},\ \bibinfo {pages} {140403} (\bibinfo {year}
  {2012})}\BibitemShut {NoStop}%
\bibitem [{\citenamefont {Fridman}\ \emph {et~al.}(2013)\citenamefont
  {Fridman}, \citenamefont {Kosmachev},\ and\ \citenamefont
  {Klevets}}]{Fridman2013}%
  \BibitemOpen
  \bibfield  {author} {\bibinfo {author} {\bibfnamefont {Yu.A.}\ \bibnamefont
  {Fridman}}, \bibinfo {author} {\bibfnamefont {O.A.}\ \bibnamefont
  {Kosmachev}}, \ and\ \bibinfo {author} {\bibfnamefont {Ph.N.}\ \bibnamefont
  {Klevets}},\ }\bibfield  {title} {\enquote {\bibinfo {title} {Spin nematic
  and orthogonal nematic states in s=1 non-heisenberg magnet},}\ }\href
  {\doibase https://doi.org/10.1016/j.jmmm.2012.08.027} {\bibfield  {journal}
  {\bibinfo  {journal} {Journal of Magnetism and Magnetic Materials}\ }\textbf
  {\bibinfo {volume} {325}},\ \bibinfo {pages} {125--129} (\bibinfo {year}
  {2013})}\BibitemShut {NoStop}%
\bibitem [{\citenamefont {Ueltschi}(2015)}]{Ueltschi2015}%
  \BibitemOpen
  \bibfield  {author} {\bibinfo {author} {\bibfnamefont {Daniel}\ \bibnamefont
  {Ueltschi}},\ }\bibfield  {title} {\enquote {\bibinfo {title}
  {Ferromagnetism, antiferromagnetism, and the curious nematic phase of $s=1$
  quantum spin systems},}\ }\href {\doibase 10.1103/PhysRevE.91.042132}
  {\bibfield  {journal} {\bibinfo  {journal} {Phys. Rev. E}\ }\textbf {\bibinfo
  {volume} {91}},\ \bibinfo {pages} {042132} (\bibinfo {year}
  {2015})}\BibitemShut {NoStop}%
\bibitem [{\citenamefont {Björnberg}\ \emph {et~al.}(2020)\citenamefont
  {Björnberg}, \citenamefont {Fröhlich},\ and\ \citenamefont
  {Ueltschi}}]{Bjornberg2020}%
  \BibitemOpen
  \bibfield  {author} {\bibinfo {author} {\bibfnamefont {Jakob~E.}\
  \bibnamefont {Björnberg}}, \bibinfo {author} {\bibfnamefont {Jürg}\
  \bibnamefont {Fröhlich}}, \ and\ \bibinfo {author} {\bibfnamefont {Daniel}\
  \bibnamefont {Ueltschi}},\ }\bibfield  {title} {\enquote {\bibinfo {title}
  {Quantum spins and random loops on the complete graph},}\ }\href {\doibase
  10.1007/s00220-019-03634-x} {\bibfield  {journal} {\bibinfo  {journal}
  {Communications in Mathematical Physics}\ }\textbf {\bibinfo {volume}
  {375}},\ \bibinfo {pages} {1629–1663} (\bibinfo {year} {2020})}\BibitemShut
  {NoStop}%
\bibitem [{\citenamefont {T{\'o}th}(1993)}]{Toth1993}%
  \BibitemOpen
  \bibfield  {author} {\bibinfo {author} {\bibfnamefont {B{\'a}lint}\
  \bibnamefont {T{\'o}th}},\ }\bibfield  {title} {\enquote {\bibinfo {title}
  {Improved lower bound on the thermodynamic pressure of the spin 1/2
  heisenberg ferromagnet},}\ }\href@noop {} {\bibfield  {journal} {\bibinfo
  {journal} {Letters in Mathematical Physics}\ }\textbf {\bibinfo {volume}
  {28}},\ \bibinfo {pages} {75--84} (\bibinfo {year} {1993})}\BibitemShut
  {NoStop}%
\bibitem [{\citenamefont {Aizenman}\ and\ \citenamefont
  {Nachtergaele}(1994)}]{Aizenman1994}%
  \BibitemOpen
  \bibfield  {author} {\bibinfo {author} {\bibfnamefont {Michael}\ \bibnamefont
  {Aizenman}}\ and\ \bibinfo {author} {\bibfnamefont {Bruno}\ \bibnamefont
  {Nachtergaele}},\ }\bibfield  {title} {\enquote {\bibinfo {title} {Geometric
  aspects of quantum spin states},}\ }\href {\doibase 10.1007/BF02108805}
  {\bibfield  {journal} {\bibinfo  {journal} {Communications in Mathematical
  Physics}\ }\textbf {\bibinfo {volume} {164}},\ \bibinfo {pages} {17--63}
  (\bibinfo {year} {1994})}\BibitemShut {NoStop}%
\bibitem [{\citenamefont {Ueltschi}(2013)}]{Ueltschi2013}%
  \BibitemOpen
  \bibfield  {author} {\bibinfo {author} {\bibfnamefont {Daniel}\ \bibnamefont
  {Ueltschi}},\ }\bibfield  {title} {\enquote {\bibinfo {title} {Random loop
  representations for quantum spin systems},}\ }\href {\doibase
  10.1063/1.4817865} {\bibfield  {journal} {\bibinfo  {journal} {Journal of
  Mathematical Physics}\ }\textbf {\bibinfo {volume} {54}},\ \bibinfo {pages}
  {083301} (\bibinfo {year} {2013})}\BibitemShut {NoStop}%
\bibitem [{\citenamefont {Goldschmidt}\ \emph {et~al.}(2011)\citenamefont
  {Goldschmidt}, \citenamefont {Ueltschi},\ and\ \citenamefont
  {Windridge}}]{Goldschmidt2011}%
  \BibitemOpen
  \bibfield  {author} {\bibinfo {author} {\bibfnamefont {Christina}\
  \bibnamefont {Goldschmidt}}, \bibinfo {author} {\bibfnamefont {Daniel}\
  \bibnamefont {Ueltschi}}, \ and\ \bibinfo {author} {\bibfnamefont {Peter}\
  \bibnamefont {Windridge}},\ }\bibfield  {title} {\enquote {\bibinfo {title}
  {Quantum {H}eisenberg models and their probabilistic representations},}\ }in\
  \href {\doibase 10.1090/conm/552/10917} {\emph {\bibinfo {booktitle} {Entropy
  and the quantum {II}}}},\ \bibinfo {series} {Contemp. Math.}, Vol.\ \bibinfo
  {volume} {552}\ (\bibinfo  {publisher} {Amer. Math. Soc., Providence, RI},\
  \bibinfo {year} {2011})\ pp.\ \bibinfo {pages} {177--224}\BibitemShut
  {NoStop}%
\bibitem [{\citenamefont {Grosskinsky}\ \emph {et~al.}(2012)\citenamefont
  {Grosskinsky}, \citenamefont {Lovisolo},\ and\ \citenamefont
  {Ueltschi}}]{Grosskinsky2012}%
  \BibitemOpen
  \bibfield  {author} {\bibinfo {author} {\bibfnamefont {Stefan}\ \bibnamefont
  {Grosskinsky}}, \bibinfo {author} {\bibfnamefont {Alexander~A.}\ \bibnamefont
  {Lovisolo}}, \ and\ \bibinfo {author} {\bibfnamefont {Daniel}\ \bibnamefont
  {Ueltschi}},\ }\bibfield  {title} {\enquote {\bibinfo {title} {Lattice
  permutations and poisson-dirichlet distribution of cycle lengths},}\ }\href
  {\doibase 10.1007/s10955-012-0450-9} {\bibfield  {journal} {\bibinfo
  {journal} {Journal of Statistical Physics}\ }\textbf {\bibinfo {volume}
  {146}},\ \bibinfo {pages} {1105--1121} (\bibinfo {year} {2012})}\BibitemShut
  {NoStop}%
\bibitem [{\citenamefont {Nahum}\ \emph {et~al.}(2013)\citenamefont {Nahum},
  \citenamefont {Chalker}, \citenamefont {Serna}, \citenamefont {Ortu\~no},\
  and\ \citenamefont {Somoza}}]{Nahum2013}%
  \BibitemOpen
  \bibfield  {author} {\bibinfo {author} {\bibfnamefont {Adam}\ \bibnamefont
  {Nahum}}, \bibinfo {author} {\bibfnamefont {J.~T.}\ \bibnamefont {Chalker}},
  \bibinfo {author} {\bibfnamefont {P.}~\bibnamefont {Serna}}, \bibinfo
  {author} {\bibfnamefont {M.}~\bibnamefont {Ortu\~no}}, \ and\ \bibinfo
  {author} {\bibfnamefont {A.~M.}\ \bibnamefont {Somoza}},\ }\bibfield  {title}
  {\enquote {\bibinfo {title} {Length distributions in loop soups},}\ }\href
  {\doibase 10.1103/PhysRevLett.111.100601} {\bibfield  {journal} {\bibinfo
  {journal} {Phys. Rev. Lett.}\ }\textbf {\bibinfo {volume} {111}},\ \bibinfo
  {pages} {100601} (\bibinfo {year} {2013})}\BibitemShut {NoStop}%
\bibitem [{\citenamefont {Ueltschi}(2017)}]{Ueltschi2017}%
  \BibitemOpen
  \bibfield  {author} {\bibinfo {author} {\bibfnamefont {Daniel}\ \bibnamefont
  {Ueltschi}},\ }\href {\doibase 10.48550/arXiv.1703.09503} {\enquote {\bibinfo
  {title} {{Universal behaviour of 3D loop soup models", in 6th Warsaw School
  of Statistical Physics, B. Cichocki, M. Napiorkowski, J. Piasecki, P.
  Szymczak eds, pp 65-100, Warsaw University Press}},}\ } (\bibinfo {year}
  {2017})\BibitemShut {NoStop}%
\bibitem [{\citenamefont {Rychkov}(2017)}]{Rychkov2017}%
  \BibitemOpen
  \bibfield  {author} {\bibinfo {author} {\bibfnamefont {Slava}\ \bibnamefont
  {Rychkov}},\ }\href {\doibase 10.1007/978-3-319-43626-5} {\emph {\bibinfo
  {title} {{EPFL lectures on conformal field theory in D $\geq$ 3
  dimensions}}}},\ SpringerBriefs in physics\ (\bibinfo  {publisher} {Springer.
  CERN},\ \bibinfo {address} {Cham. Geneva},\ \bibinfo {year}
  {2017})\BibitemShut {NoStop}%
\bibitem [{\citenamefont {Pelissetto}\ and\ \citenamefont
  {Vicari}(2019)}]{Pelissetto2019}%
  \BibitemOpen
  \bibfield  {author} {\bibinfo {author} {\bibfnamefont {Andrea}\ \bibnamefont
  {Pelissetto}}\ and\ \bibinfo {author} {\bibfnamefont {Ettore}\ \bibnamefont
  {Vicari}},\ }\bibfield  {title} {\enquote {\bibinfo {title}
  {Three-dimensional ferromagnetic ${\mathrm{cp}}^{N\ensuremath{-}1}$
  models},}\ }\href {\doibase 10.1103/PhysRevE.100.022122} {\bibfield
  {journal} {\bibinfo  {journal} {Phys. Rev. E}\ }\textbf {\bibinfo {volume}
  {100}},\ \bibinfo {pages} {022122} (\bibinfo {year} {2019})}\BibitemShut
  {NoStop}%
\bibitem [{\citenamefont {Read}\ and\ \citenamefont
  {Sachdev}(1990)}]{Read1990}%
  \BibitemOpen
  \bibfield  {author} {\bibinfo {author} {\bibfnamefont {N.}~\bibnamefont
  {Read}}\ and\ \bibinfo {author} {\bibfnamefont {Subir}\ \bibnamefont
  {Sachdev}},\ }\bibfield  {title} {\enquote {\bibinfo {title} {Spin-peierls,
  valence-bond solid, and n\'eel ground states of low-dimensional quantum
  antiferromagnets},}\ }\href {\doibase 10.1103/PhysRevB.42.4568} {\bibfield
  {journal} {\bibinfo  {journal} {Phys. Rev. B}\ }\textbf {\bibinfo {volume}
  {42}},\ \bibinfo {pages} {4568--4589} (\bibinfo {year} {1990})}\BibitemShut
  {NoStop}%
\bibitem [{\citenamefont {Caci}\ \emph {et~al.}(2025)\citenamefont {Caci},
  \citenamefont {Chudy}, \citenamefont {Mendez~Mariscal}, \citenamefont
  {Ueltschi},\ and\ \citenamefont {Wessel}}]{Caci2025_data}%
  \BibitemOpen
  \bibfield  {author} {\bibinfo {author} {\bibfnamefont {Nils}\ \bibnamefont
  {Caci}}, \bibinfo {author} {\bibfnamefont {Dominik}\ \bibnamefont {Chudy}},
  \bibinfo {author} {\bibfnamefont {Pablo~Daniel}\ \bibnamefont
  {Mendez~Mariscal}}, \bibinfo {author} {\bibfnamefont {Daniel}\ \bibnamefont
  {Ueltschi}}, \ and\ \bibinfo {author} {\bibfnamefont {Stefan}\ \bibnamefont
  {Wessel}},\ }\href {\doibase 10.5281/zenodo.15226306} {\enquote {\bibinfo
  {title} {Figure data to "symmetry breaking and thermal phase transition of
  the spin-1 quantum magnet with su(3) symmetry on the simple cubic
  lattice"},}\ } (\bibinfo {year} {2025})\BibitemShut {NoStop}%
\bibitem [{\citenamefont {Ruelle}(1969)}]{Ruelle1969}%
  \BibitemOpen
  \bibfield  {author} {\bibinfo {author} {\bibfnamefont {David}\ \bibnamefont
  {Ruelle}},\ }\href@noop {} {\emph {\bibinfo {title} {{S}tatistical
  {M}echanics}}}\ (\bibinfo  {publisher} {W.A. Benjamin},\ \bibinfo {year}
  {1969})\BibitemShut {NoStop}%
\bibitem [{\citenamefont {Israel}(2016)}]{Israel_2016}%
  \BibitemOpen
  \bibfield  {author} {\bibinfo {author} {\bibfnamefont {Robert~B.}\
  \bibnamefont {Israel}},\ }\href@noop {} {\emph {\bibinfo {title} {Convexity
  in the theory of lattice gases}}}\ (\bibinfo  {publisher} {Princeton
  University Press},\ \bibinfo {year} {2016})\BibitemShut {NoStop}%
\bibitem [{\citenamefont {Itzykson}\ and\ \citenamefont
  {Zuber}(1980)}]{Itzykson1980}%
  \BibitemOpen
  \bibfield  {author} {\bibinfo {author} {\bibfnamefont {C.}~\bibnamefont
  {Itzykson}}\ and\ \bibinfo {author} {\bibfnamefont {J.-B.}\ \bibnamefont
  {Zuber}},\ }\bibfield  {title} {\enquote {\bibinfo {title} {The planar
  approximation. ii},}\ }\href {\doibase 10.1063/1.524438} {\bibfield
  {journal} {\bibinfo  {journal} {Journal of Mathematical Physics}\ }\textbf
  {\bibinfo {volume} {21}},\ \bibinfo {pages} {411–421} (\bibinfo {year}
  {1980})}\BibitemShut {NoStop}%
\bibitem [{\citenamefont {Sandvik}\ and\ \citenamefont
  {Kurkij\"arvi}(1991)}]{Sandvik1991}%
  \BibitemOpen
  \bibfield  {author} {\bibinfo {author} {\bibfnamefont {Anders~W.}\
  \bibnamefont {Sandvik}}\ and\ \bibinfo {author} {\bibfnamefont {Juhani}\
  \bibnamefont {Kurkij\"arvi}},\ }\bibfield  {title} {\enquote {\bibinfo
  {title} {Quantum {M}onte {C}arlo simulation method for spin systems},}\
  }\href {\doibase 10.1103/PhysRevB.43.5950} {\bibfield  {journal} {\bibinfo
  {journal} {Phys. Rev. B}\ }\textbf {\bibinfo {volume} {43}},\ \bibinfo
  {pages} {5950--5961} (\bibinfo {year} {1991})}\BibitemShut {NoStop}%
\bibitem [{\citenamefont {Sandvik}(1999)}]{Sandvik1999}%
  \BibitemOpen
  \bibfield  {author} {\bibinfo {author} {\bibfnamefont {Anders~W.}\
  \bibnamefont {Sandvik}},\ }\bibfield  {title} {\enquote {\bibinfo {title}
  {Stochastic series expansion method with operator-loop update},}\ }\href
  {\doibase 10.1103/PhysRevB.59.R14157} {\bibfield  {journal} {\bibinfo
  {journal} {Phys. Rev. B}\ }\textbf {\bibinfo {volume} {59}},\ \bibinfo
  {pages} {R14157--R14160} (\bibinfo {year} {1999})}\BibitemShut {NoStop}%
\bibitem [{\citenamefont {Sylju\aa{}sen}\ and\ \citenamefont
  {Sandvik}(2002)}]{Syljuasen2002}%
  \BibitemOpen
  \bibfield  {author} {\bibinfo {author} {\bibfnamefont {Olav~F.}\ \bibnamefont
  {Sylju\aa{}sen}}\ and\ \bibinfo {author} {\bibfnamefont {Anders~W.}\
  \bibnamefont {Sandvik}},\ }\bibfield  {title} {\enquote {\bibinfo {title}
  {Quantum monte carlo with directed loops},}\ }\href {\doibase
  10.1103/PhysRevE.66.046701} {\bibfield  {journal} {\bibinfo  {journal} {Phys.
  Rev. E}\ }\textbf {\bibinfo {volume} {66}},\ \bibinfo {pages} {046701}
  (\bibinfo {year} {2002})}\BibitemShut {NoStop}%
\bibitem [{\citenamefont {Alet}\ \emph {et~al.}(2005)\citenamefont {Alet},
  \citenamefont {Wessel},\ and\ \citenamefont {Troyer}}]{Alet2005}%
  \BibitemOpen
  \bibfield  {author} {\bibinfo {author} {\bibfnamefont {Fabien}\ \bibnamefont
  {Alet}}, \bibinfo {author} {\bibfnamefont {Stefan}\ \bibnamefont {Wessel}}, \
  and\ \bibinfo {author} {\bibfnamefont {Matthias}\ \bibnamefont {Troyer}},\
  }\bibfield  {title} {\enquote {\bibinfo {title} {Generalized directed loop
  method for quantum monte carlo simulations},}\ }\href {\doibase
  10.1103/PhysRevE.71.036706} {\bibfield  {journal} {\bibinfo  {journal} {Phys.
  Rev. E}\ }\textbf {\bibinfo {volume} {71}},\ \bibinfo {pages} {036706}
  (\bibinfo {year} {2005})}\BibitemShut {NoStop}%
\bibitem [{\citenamefont {Nagata}\ \emph {et~al.}(2001)\citenamefont {Nagata},
  \citenamefont {Žukovič},\ and\ \citenamefont {Idogaki}}]{Nagata2001}%
  \BibitemOpen
  \bibfield  {author} {\bibinfo {author} {\bibfnamefont {H.}~\bibnamefont
  {Nagata}}, \bibinfo {author} {\bibfnamefont {M.}~\bibnamefont {Žukovič}}, \
  and\ \bibinfo {author} {\bibfnamefont {T.}~\bibnamefont {Idogaki}},\
  }\bibfield  {title} {\enquote {\bibinfo {title} {Monte carlo simulation of
  the three-dimensional xy model with bilinear–biquadratic exchange
  interaction},}\ }\href {\doibase
  https://doi.org/10.1016/S0304-8853(01)00392-4} {\bibfield  {journal}
  {\bibinfo  {journal} {Journal of Magnetism and Magnetic Materials}\ }\textbf
  {\bibinfo {volume} {234}},\ \bibinfo {pages} {320--330} (\bibinfo {year}
  {2001})}\BibitemShut {NoStop}%
\bibitem [{\citenamefont {Campostrini}\ \emph {et~al.}(2001)\citenamefont
  {Campostrini}, \citenamefont {Hasenbusch}, \citenamefont {Pelissetto},
  \citenamefont {Rossi},\ and\ \citenamefont {Vicari}}]{Campostrini2001}%
  \BibitemOpen
  \bibfield  {author} {\bibinfo {author} {\bibfnamefont {Massimo}\ \bibnamefont
  {Campostrini}}, \bibinfo {author} {\bibfnamefont {Martin}\ \bibnamefont
  {Hasenbusch}}, \bibinfo {author} {\bibfnamefont {Andrea}\ \bibnamefont
  {Pelissetto}}, \bibinfo {author} {\bibfnamefont {Paolo}\ \bibnamefont
  {Rossi}}, \ and\ \bibinfo {author} {\bibfnamefont {Ettore}\ \bibnamefont
  {Vicari}},\ }\bibfield  {title} {\enquote {\bibinfo {title} {Critical
  behavior of the three-dimensional $\mathrm{XY}$ universality class},}\ }\href
  {\doibase 10.1103/PhysRevB.63.214503} {\bibfield  {journal} {\bibinfo
  {journal} {Phys. Rev. B}\ }\textbf {\bibinfo {volume} {63}},\ \bibinfo
  {pages} {214503} (\bibinfo {year} {2001})}\BibitemShut {NoStop}%
\bibitem [{\citenamefont {Privman}\ and\ \citenamefont
  {Fisher}(1988)}]{Privman1988}%
  \BibitemOpen
  \bibfield  {author} {\bibinfo {author} {\bibfnamefont {Vladimir}\
  \bibnamefont {Privman}}\ and\ \bibinfo {author} {\bibfnamefont {Michael~E.}\
  \bibnamefont {Fisher}},\ }\bibfield  {title} {\enquote {\bibinfo {title}
  {Finite-size effects at first-order transitions},}\ }in\ \href {\doibase
  10.1016/b978-0-444-87109-1.50017-0} {\emph {\bibinfo {booktitle} {Finite-Size
  Scaling}}}\ (\bibinfo  {publisher} {Elsevier},\ \bibinfo {year} {1988})\ pp.\
  \bibinfo {pages} {149--181}\BibitemShut {NoStop}%
\bibitem [{\citenamefont {Janke}(1993)}]{Janke1993}%
  \BibitemOpen
  \bibfield  {author} {\bibinfo {author} {\bibfnamefont {W.}~\bibnamefont
  {Janke}},\ }\bibfield  {title} {\enquote {\bibinfo {title} {Accurate
  first-order transition points from finite-size data without power-law
  corrections},}\ }\href {\doibase 10.1103/PhysRevB.47.14757} {\bibfield
  {journal} {\bibinfo  {journal} {Phys. Rev. B}\ }\textbf {\bibinfo {volume}
  {47}},\ \bibinfo {pages} {14757--14770} (\bibinfo {year} {1993})}\BibitemShut
  {NoStop}%
\bibitem [{\citenamefont {Lee}\ and\ \citenamefont
  {Kosterlitz}(1990)}]{Lee1990}%
  \BibitemOpen
  \bibfield  {author} {\bibinfo {author} {\bibfnamefont {Jooyoung}\
  \bibnamefont {Lee}}\ and\ \bibinfo {author} {\bibfnamefont {J.~M.}\
  \bibnamefont {Kosterlitz}},\ }\bibfield  {title} {\enquote {\bibinfo {title}
  {New numerical method to study phase transitions},}\ }\href {\doibase
  10.1103/PhysRevLett.65.137} {\bibfield  {journal} {\bibinfo  {journal} {Phys.
  Rev. Lett.}\ }\textbf {\bibinfo {volume} {65}},\ \bibinfo {pages} {137--140}
  (\bibinfo {year} {1990})}\BibitemShut {NoStop}%
\bibitem [{\citenamefont {Lee}\ and\ \citenamefont
  {Kosterlitz}(1991)}]{Lee1991}%
  \BibitemOpen
  \bibfield  {author} {\bibinfo {author} {\bibfnamefont {Jooyoung}\
  \bibnamefont {Lee}}\ and\ \bibinfo {author} {\bibfnamefont {J.~M.}\
  \bibnamefont {Kosterlitz}},\ }\bibfield  {title} {\enquote {\bibinfo {title}
  {Finite-size scaling and monte carlo simulations of first-order phase
  transitions},}\ }\href {\doibase 10.1103/PhysRevB.43.3265} {\bibfield
  {journal} {\bibinfo  {journal} {Phys. Rev. B}\ }\textbf {\bibinfo {volume}
  {43}},\ \bibinfo {pages} {3265--3277} (\bibinfo {year} {1991})}\BibitemShut
  {NoStop}%
\bibitem [{\citenamefont {Wu}(1982)}]{Wu1982}%
  \BibitemOpen
  \bibfield  {author} {\bibinfo {author} {\bibfnamefont {F.~Y.}\ \bibnamefont
  {Wu}},\ }\bibfield  {title} {\enquote {\bibinfo {title} {The potts model},}\
  }\href {\doibase 10.1103/RevModPhys.54.235} {\bibfield  {journal} {\bibinfo
  {journal} {Rev. Mod. Phys.}\ }\textbf {\bibinfo {volume} {54}},\ \bibinfo
  {pages} {235--268} (\bibinfo {year} {1982})}\BibitemShut {NoStop}%
\bibitem [{\citenamefont {Buddenoir}\ and\ \citenamefont
  {Wallon}(1993)}]{Buddenoir1993}%
  \BibitemOpen
  \bibfield  {author} {\bibinfo {author} {\bibfnamefont {E}~\bibnamefont
  {Buddenoir}}\ and\ \bibinfo {author} {\bibfnamefont {S}~\bibnamefont
  {Wallon}},\ }\bibfield  {title} {\enquote {\bibinfo {title} {The correlation
  length of the potts model at the first-order transition point},}\ }\href
  {\doibase 10.1088/0305-4470/26/13/009} {\bibfield  {journal} {\bibinfo
  {journal} {Journal of Physics A: Mathematical and General}\ }\textbf
  {\bibinfo {volume} {26}},\ \bibinfo {pages} {3045} (\bibinfo {year}
  {1993})}\BibitemShut {NoStop}%
\bibitem [{\citenamefont {Kent-Dobias}\ and\ \citenamefont
  {Sethna}(2018)}]{Kent-Dobias2018}%
  \BibitemOpen
  \bibfield  {author} {\bibinfo {author} {\bibfnamefont {Jaron}\ \bibnamefont
  {Kent-Dobias}}\ and\ \bibinfo {author} {\bibfnamefont {James~P.}\
  \bibnamefont {Sethna}},\ }\bibfield  {title} {\enquote {\bibinfo {title}
  {Cluster representations and the wolff algorithm in arbitrary external
  fields},}\ }\href {\doibase 10.1103/PhysRevE.98.063306} {\bibfield  {journal}
  {\bibinfo  {journal} {Phys. Rev. E}\ }\textbf {\bibinfo {volume} {98}},\
  \bibinfo {pages} {063306} (\bibinfo {year} {2018})}\BibitemShut {NoStop}%
\end{thebibliography}%


\end{document}